\documentclass{aa}  
\usepackage{fixltx2e}
\usepackage[T1]{fontenc}
\usepackage{graphicx}
\usepackage{lmodern}
\usepackage{listings}
\usepackage{multirow}

\newcommand{\inp}{_\mathrm{input}}
\newcommand{\out}{_\mathrm{output}}

\newcommand{\e}[1]{\times 10^{#1}}
\newcommand{\SN}{\mathrm{S}/\mathrm{N}}

\newcommand{\sal}{\sigma_{\!\mathrm{A\!L}}}
\newcommand{\msun}{M_\odot}

\newcommand{\dchisq}{\Delta\chi^2}

\newcommand{\stan}{\texttt{Stan}}
\newcommand{\Dic}{\Delta \mathrm{IC}}

\begin{document}

   \title{Astrometry and exoplanets in the Gaia era:\\ a
     Bayesian approach to detection and parameter recovery}
   \titlerunning{Astrometry and exoplanets in the Gaia era: a Bayesian approach}

   \author{P. Ranalli       \inst{1,2}
         \and
         D. Hobbs           \inst{1}
         \and
         L. Lindegren       \inst{1}
          }

   \institute{
     Lund Observatory, Box 43, 22100 Lund, Sweden \\
     \email{piero@astro.lu.se}
   \and
     INAF -- Osservatorio Astronomico di Bologna,
     via Gobetti 93/3, 40129 Bologna, Italy
   }

\abstract{
The Gaia mission is expected to make a significant contribution to the
knowledge of exoplanet systems, both in terms of their number and of
their physical properties.
We develop Bayesian methods and detection criteria for
orbital fitting, and revise the detectability of exoplanets in
light of the in-flight properties of Gaia.
Limiting ourselves to one-planet systems as a first step of the
development, we simulate Gaia data for exoplanet systems over a grid
of S/N,  orbital period, and eccentricity. The simulations are then fit
using Markov chain Monte Carlo methods.
We investigate the detection rate
according to three information criteria and the $\dchisq$. For the
$\dchisq$, the effective number of degrees of freedom
depends on the mission length.
We find that the choice of the Markov chain starting point can affect
the quality of the results; we therefore consider two limit
possibilities: an ideal case, and a very simple method that
finds the starting point assuming circular orbits.
We use 6644 and 4402 simulations to assess the fraction of false
positive detections in a 5~yr and in a 10~yr mission,
respectively; and 4968 and 4706 simulations to assess the detection
rate and how the parameters are recovered.
Using Jeffreys' scale of evidence, the fraction of false
positives passing a strong evidence criterion is $\lesssim 0.2\%$
(0.6\%) when considering a 5~yr (10~yr) mission and
using the Akaike information criterion or the Watanabe--Akaike
information criterion, and
$<0.02\%$ ($<0.06\%$) when using the Bayesian information
criterion. We find that there is a 50\% chance of detecting a planet
with a minimum S/N=2.3 (1.7).
This sets the maximum
distance to which a planet is detectable to $\sim 70$~pc and
$\sim 3.5$~pc for a Jupiter-mass and Neptune-mass planet, respectively, assuming
a 10~yr mission, a 4~au semi-major axis, and a 1~$M_\sun$
star.  
We show the distribution of the accuracy and precision with which
orbital parameters are recovered.
The period is the orbital parameter
that can be determined with the best accuracy, with a median relative
difference between input and output periods of $4.2\%$ (2.9\%)
assuming a 5~yr (10~yr) mission. The median
accuracy of the
semi-major axis of the orbit   can be recovered with a median relative
error of $7\%$ (6\%). The eccentricity can also be recovered with a
median absolute accuracy of $0.07$ (0.06).
}


   \maketitle

\section{Introduction}
\label{sec:intro}

The detection of exoplanets with astrometric techniques is simple in
principle. Although an exoplanet is usually too faint to be observed
with current instruments,  its existence can be inferred from the
motion of its host star around the centre of mass of the star-planet
system. In the past, several claims of detections have been made and
later disproved starting from \citet{jacob1855} (see
\citet{quirrenbach-in-seager} for a brief historical account). The
difficulty lies in the small magnitude of the host star astrometric
shift, which even in the most favourable cases is $\lesssim 1$ mas
(see e.g.\ Sect.~\ref{sec:SNdetectability}).

The astrometric satellite Hipparcos \citep{hipparcos1997} could
measure the along-scan position of a star with an error of a few
mas. Several studies tried to derive orbital constraints for candidate
planets suggested by radial velocity searches
\citep{mazeh1999,zucker2000,gatewood2001,han2001}, but fitting
orbital models to Hipparcos data hardly improved the astrometric fit
and led instead to spurious features \citep{pourbaix2001}.

The Gaia mission \citep{gaia2016} offers a $\sim 30$-fold improvement
in astrometric precision with respect to Hipparcos, and can therefore
finally allow the astrometric detection of planets to become
routine. The Gaia capabilities for planet detection were soon
  recognised \citep{bernstein1995,casertano1995}, and explored in depth by
\citet[hereafter C08]{casertano2008} and later revised by
\citet[hereafter P14]{perryman2014}. The number of detectable planets
was estimated by P14 to be $\sim 21,000$ for a 5 yr mission, after
considering a model of the Galaxy population and fitting planetary
orbits with the least-squares method.

In this paper we consider the application of Bayesian methods to the
problem of planetary orbit fitting. We focus on two aspects of
Bayesian inference: the derivation of joint errors for all parameters
of interest, and the comparison of models. The first builds on the
ability of Markov chain Monte Carlo (MCMC) methods to track errors
beyond the regime in which linear approximations are acceptable. The
second is especially important in light of the difficulties that arise
for calculating goodness of fit for non-linear models. Our discussion
also includes a traditional metric, $\dchisq$, that can be used to
achieve results similar to those obtained using information criteria.
Bayesian methods are already in use for the Gaia data processing, and
thus we expect our results to better anticipate the outcome of future
Gaia data releases.

Similarly to C08 and at variance with P14, we focus more on the
detectability as a function of the orbital characteristics than on
the number of detectable planets in the solar neighbourhood. While C08
was published five years before the launch of Gaia and therefore 
reflects pre-launch expectations, in this paper we use the current
knowledge of the satellite to reassess what fraction of planets can be
detected and how well their orbital parameters can be recovered.

In this paper we focus on developing methods, and so we only consider
single-planet systems.  While this may be a reasonable approximation
in some cases, real-world planetary systems can have more than one
planet with a non-negligible signal-to-noise ratio (S/N), in which case
the results discussed here may not apply.  Double-planet and more
complicated systems will be considered in a follow-up paper.

The structure of this paper is as follows. In
Sect.~\ref{sec:astrometry} we describe our adopted formalism for the
description of Keplerian orbits in an astrometry setting. In
Sect.~\ref{sec:SNdetectability} we consider the expected S/N
for a one-planet system. In Sect.~\ref{sec:simulations} we
describe our simulations. In Sect.~\ref{sec:bayes} we consider
Bayesian methods and detection metrics. In
Sect.~\ref{sec:falsepositives} we estimate the fraction of false
positive detections. In Sect.~\ref{sec:detectionrate} we show the
detection rate. In Sect.~\ref{sec:param-recovery} we discuss
how the orbital parameters are recovered.  In
Sect.~\ref{sec:discussion} we discuss our results, and include an
assessment of the average accuracy and precision which a
parameter can be recovered. Finally, in Sect.~\ref{sec:conclusions}, we present our
conclusions.

\section{Astrometry and Keplerian orbits}
\label{sec:astrometry}

\subsection{Formalism}
\label{sec:orbits}

\begin{figure}
  \centering
  \resizebox{\hsize}{!}{\includegraphics[width=\columnwidth,bb=88 129 373 241,clip]{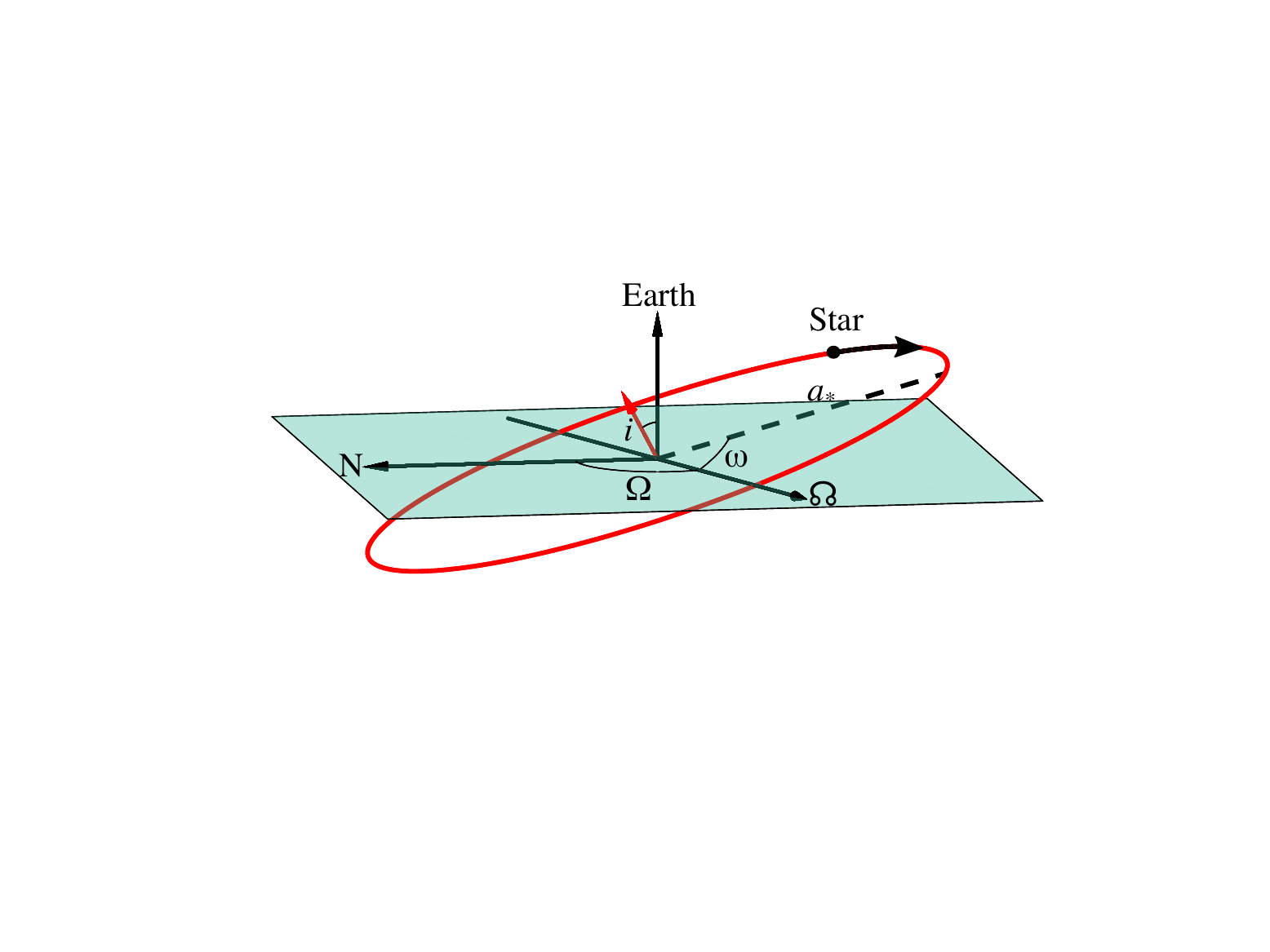}}
  \caption{Illustration of the four geometrical elements: the
    semi-major axis $a_*$, the longitude of the ascending node $\Omega$
    (also known as receding node), the inclination $i$, and
    the argument of periapsis $\omega$. The light cyan polygon
    represents the sky plane in a perspective view; the reference
    direction is taken towards the north (N). The line of nodes, shown
    by the black arrow pointing towards the ascending node,
    is the intersection between the sky and the orbit
    planes. The orbit is shown as the red ellipse.
    The small red arrow is perpendicular to the true plane
    and shows the inclination.}
  \label{fig:orbitalelements}
\end{figure}

An exoplanet orbit can be determined in the same way as that of
astrometric binary stars, where only one element of the binary system
is visible. Here we summarise  the method; secondary references
include \citet{binnendijk},
\citet{taff1985}, \citet{quirrenbach-in-seager}, and
\citet{perryman2011book}.

Let us consider a system made of a star and one planet.  In an
inertial reference frame, both bodies follow elliptical orbits around
the system's centre of mass.  The semi-major axis of the star's orbit is
\begin{equation}
  \label{eq:a_star}
  a_*=a_p M_p/M_*
,\end{equation}
where $a_p$ is the semi-major axis of the planet orbit, $M_p$ is the
planet mass, and $M_*$ is the stellar mass. The astrometric
signature\footnote{From the Greek
  $\upsilon\pi o\gamma\rho\alpha\phi\acute{\eta}$ (signature).}
$\upsilon$ is the angular size of the stellar orbit
\begin{equation}
  \label{eq:astrom-signature}
  \upsilon           = \left( \frac{M_p}{M_*} \right)
                \left( \frac{a_p}{\mathrm{AU}} \right)
                \left( \frac{d}{\mathrm{pc}} \right)^{-1}
                \quad \mathrm{arcsec}
,\end{equation}
where $d$ is the distance from the observer to the planetary system.

The orbital elements define the orbit on the true plane. Three of them
are called dynamical elements as they define the motion in the orbit:
the period $P$, the time of periastron passage $T_p$, and the
eccentricity $e$. Four geometrical elements describe the size of the
orbit and its orientation with respect to the plane of the sky: the
semi-major axis $a_*$, the longitude of the ascending node $\Omega$,
the argument of periapsis $\omega$, and the inclination
$i$ (see Fig.~\ref{fig:orbitalelements}).

The instantaneous position on the sky of a star at time $t$ is
determined by a fixed displacement ($\Delta \alpha_0$,
$\Delta \delta_0$) from its nominal position ($\alpha_0,\delta_0$) at
a reference time $t_0$, the proper motion vector\footnote{ The proper
  motion component in the right ascension includes the $\cos \delta$
  factor, and corresponds to the $\mu_{\alpha^*}$ in Appendix A of
  \citet{michalik2014}. } ($\mu_{\alpha}$, $\mu_\delta$), the parallax
$\varpi$, and the orbital elements. In Cartesian coordinates
lying on a plane tangent to the sky and directed along right ascension
and declination, respectively \citep{michalik2014}, the position can be written as
\begin{alignat}{6}
  \label{eq:stellarmotion1}
  &\Delta \alpha(t) &= \Delta\alpha_0 &+ \Pi_{\alpha} \varpi &+ (t-t_0)\mu_{a}
           &+ BX(t) &+ GY(t) ,\\
  \label{eq:stellarmotion2}
  &\Delta \delta(t)   &= \Delta\delta_0   &+ \Pi_\delta\, \varpi  &+ (t-t_0)\mu_\delta
           &+ AX(t) &+ FY(t),
\end{alignat}
where $\Pi_{\alpha}$ and $\Pi_\delta$ are the components of the
parallax direction, $X$ and $Y$ are the displacement per unit
amplitude due to planetary motion in the true plane projected onto
the sky through the Thiele--Innes (TI) constants, also called natural
elements $A$, $B$, $F$, $G$
\citep{thiele1883,vandenbos1926-innes,vandenbos1932-innes}:
\begin{alignat}{3}
  \label{eq:TI}
  &A = \upsilon\, (  & \cos \omega \cos \Omega &- \sin \omega \sin \Omega \cos i ) \\
  &B = \upsilon\, (  & \cos \omega \sin \Omega &+ \sin \omega \cos \Omega \cos i ) \\
  &F = \upsilon\, ( -& \sin \omega \cos \Omega &- \cos \omega \sin \Omega \cos i ) \\
  &G = \upsilon\, ( -& \sin \omega \sin \Omega &+ \cos \omega \cos \Omega \cos i ) 
\end{alignat}

In the true plane the coordinates $X$ and $Y$ depend on the eccentric
anomaly $E$ and on the eccentricity $e$:
\begin{align}
  \label{eq:XY}
  X(t) &= \cos E(t) - e \\
  Y(t) &= \sqrt{1-e^2}\, \sin E(t) \quad.
\end{align}
The eccentric anomaly is obtained from Kepler's equation
\begin{equation}
  \label{eq:kepler}
  E(t) = M(t) + e \sin E(t)
,\end{equation}
where $M$ is the mean anomaly:
\begin{equation}
  \label{eq:mean-anomaly}
  M(t) = \frac{360^\circ}{P} (t-T_p) \quad.
\end{equation}

Within the TI formalism, the determination of the orbital elements
from astrometric observations is a non-linear problem in $P$, $e$, and
$T_p$, and linear in all other variables.

\subsection{Degeneracies}
\label{sec:degeneracies}

In some cases, some parameters can become degenerate. For instance, in
the case of circular orbits, the argument of periapsis $\omega$ loses
its geometrical meaning as direction of the semi-major axis of the
orbital ellipse, and it becomes degenerate with the periapsis transit
time $T_p$.

A second case is that of face-on orbits, where $\cos i=\pm1$. Here,
the true plane of the orbit coincides with the sky plane, so that both
the longitude of the ascending node $\Omega$ and the argument of
periapsis $\omega$ are measured on the same plane. The intersection of
the two planes, which defines the ascending node, becomes
undefined. If Eqs.~(\ref{eq:TI}--8) are left unchanged, then the sum
$\Omega+\omega$ identifies the reference direction from which the time
of periastron passage $T_p$ is measured, but $\Omega$ and $\omega$
themselves become undefined. A possible change to the model could be
to use a different definition for $\omega$  (e.g. measure it from the
north direction).

A third case happens because astrometric data cannot distinguish the
ascending node from the descending one. In principle, both the
longitude of the ascending node $\Omega$ and the argument of periapsis
$\omega$ vary between $-180^\circ$ and $180^\circ$. However, the same
projected orbit can be obtained by adding $180^\circ$ to both $\Omega$
and $\omega$, i.e.\ by exchanging the ascending node for the
descending node and the periapsis for the apoapsis. It is customary
when analysing astrometric data to restrict the allowed range of
$\Omega$ to the $[0^\circ,180^\circ]$ interval
\citep{binnendijk,perryman2014}.  This degeneracy could be broken if
one were to jointly analyse astrometric and radial-velocity data.

The above degeneracies are observed in our analysis and the affected
systems will be identified in Sect.~\ref{sec:bayes-results-1pl}.

\section{Detectability as a function of the signal-to-noise ratio}
\label{sec:SNdetectability}

\begin{figure}
  \centering
  \resizebox{\hsize}{!}{\includegraphics[width=\columnwidth]{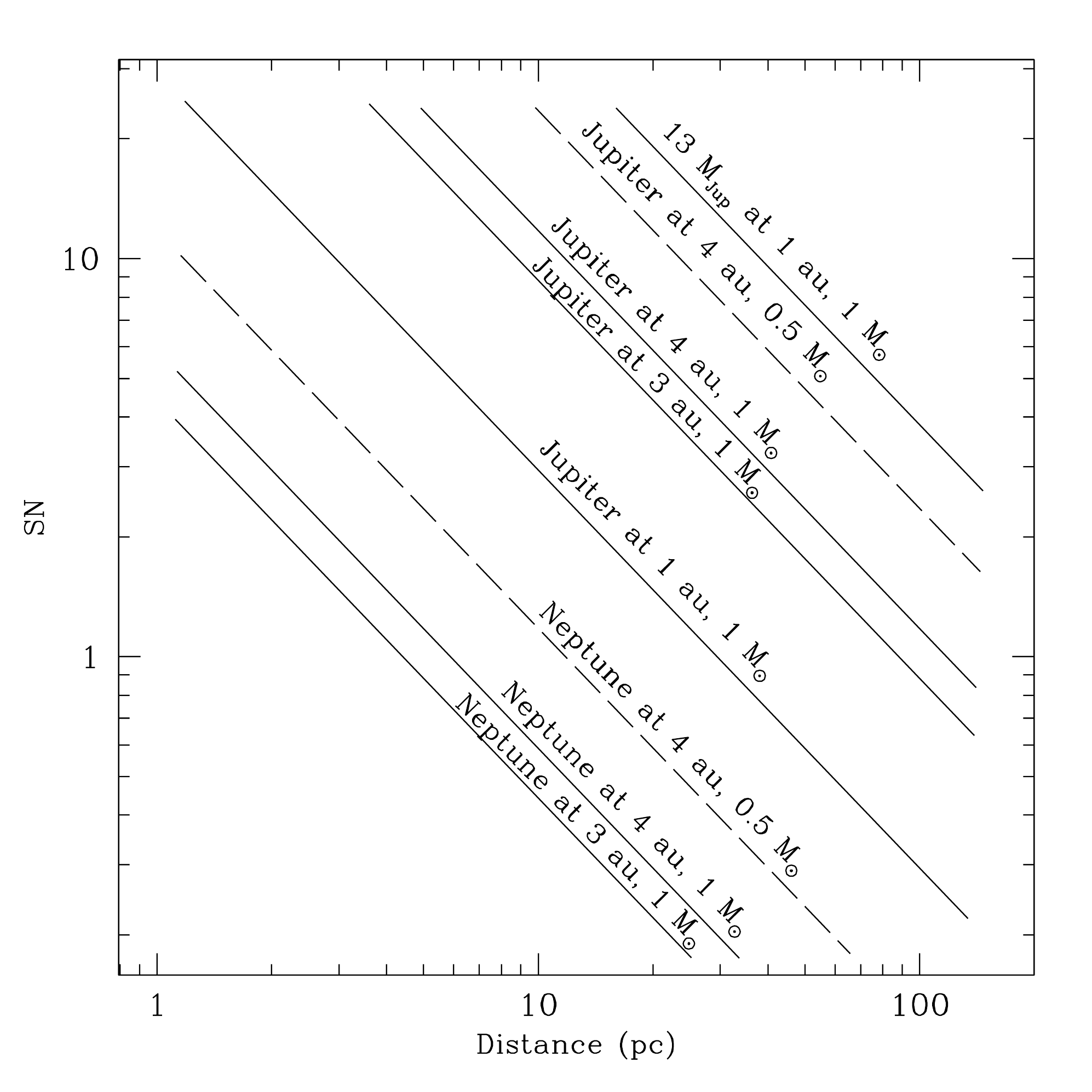}}
  \caption{Expected signal-to-noise ratio (S/N) for different
    planet masses as a function of distance. The lines show the cases
    for 13 Jupiter masses, 1 Jupiter mass, and 1 Neptune mass; with
    circular orbits at 1, 3, and 4 au (it is only possible to observe
    a 4 au orbit if the mission length is $\sim 10$~yr; see
    Sect.~\ref{sec:param-recovery}). A $1\, M_\odot$ star is
    assumed for the solid lines, while a $0.5\, M_\odot$
    star is assumed for the dashed lines.}
  \label{fig:sn-vs-d}
\end{figure}

Previous works by C08 and P14 have shown that the main parameters,
upon which the detectability of a planet depends, are the period $P$
and the signal-to-noise ratio
\begin{equation}
  \label{eq:SN}
  \SN = \frac{\upsilon}{\sal}
,\end{equation}
where $\upsilon$ is the astrometric signature and $\sal$ is the single-scan
uncertainty on our observable, the shift in the along-scan (AL) field angle $\eta$
(defined in Sect.~\ref{sec:field-angles}).

Detectability may also depend on the period because orbit sampling
becomes difficult or incomplete either if $P$ is shorter than the 
average time between two Gaia scans, or if $P$ is longer than the
mission length. This is explored in Sect.~\ref{sec:detectionrate}.

Any value of S/N can be mapped, using Eq.~(\ref{eq:astrom-signature}),
to a subspace of $M_p$, $M_*$, $a_p$, $d$, and of the stellar
magnitude $V$, which would produce the same simulated data. This is
illustrated for a few example cases in Fig.~\ref{fig:sn-vs-d}, where
planets of different masses (1 Neptune mass, 1 Jupiter mass,
  and 13 Jupiter masses\footnote{A mass of 13 $M_\mathrm{Jup}$ is the
    approximate limit mass for deuterium burning, and it is usually
    taken as the upper limit for planetary masses
    \citep{boss2007-IAUplanetdefinition}.}) are assumed to orbit a
bright star at different semi-major axes and distances. Both C08 and
P14 suggested that a $\SN\gtrsim 3$ is needed to detect a planet;
therefore, Fig.~\ref{fig:sn-vs-d} also shows an approximate
estimate of the distance from Earth up to which a planet can be
detected.

Therefore, in the following we will parameterise our simulations over
a grid of $P$ and S/N.  Only when relevant do we trace the S/N back to
the parameters on which it depends.

\section{Simulated Gaia observations}
\label{sec:simulations}

\subsection{Field angles in the Gaia field of view}
\label{sec:field-angles}

The Gaia spacecraft scans the sky along great circles whose
orientation changes in time, following a complex pattern which we
refer to as the Gaia scanning law. For each star transiting any of the
two fields of view, the time-dependent position in the AL direction
is measured on the sky mapper charge-coupled devices (CCD) and on each
of the nine astrometric field CCDs. The formal uncertainty $\sal$
results from combining all ten measurements, and it depends on the
stellar magnitude, with $\sal=0.034$ mas for $V\le 14$ for a single
scan, and larger for fainter stars
\citep{debruijne2012,debruijne2014}. A coarser determination of the
position in the across-scan (AC) direction is also done, but issues of
pixel size and binning make AC measurements less useful for
astrometry.  Gaia returns on average 70 times to each star
during its 5~yr mission \citep{debruijne2010}, and each time the
transit in the field-of-view measurement  occurs with a different
orientation, thus the stellar position can be pinpointed with AL
information only.

If $\theta(t)$ is the position angle of the Gaia field of view, the
shift in the AL field angle $\eta$ is given by
\begin{equation}
  \label{eq:eta-field-angle}
  \Delta \eta(t) = \Delta \alpha \cos \theta(t) + \Delta \delta \sin \theta(t)
\end{equation}
and Eqs.~(\ref{eq:stellarmotion1}, \ref{eq:stellarmotion2}) can be
expressed in terms of AL displacement:
\begin{align}
  \label{eq:eta-astrometric-keplerian}
  \notag
  \Delta \eta(t) &= \Delta \alpha_0   \cos \theta(t)
                  + \Delta \delta_0     \sin \theta(t) \\
         \notag  &+ \Pi_{\alpha}\, \varpi \cos \theta(t)
                  + \Pi_\delta\, \varpi  \sin \theta(t)  \\ 
                 &+ (t-t_0) \mu_{\alpha} \cos \theta(t)
                  + (t-t_0) \mu_\delta  \sin \theta(t) \\
         \notag  &+ \left(B X(t) + G Y(t)\right) \cos \theta(t) \\
         \notag  &+  \left(A X(t) + F Y(t)\right) \sin \theta(t) \quad .
\end{align}

\subsection{Single-planet systems: simulations over a grid of parameters}
\label{sec:single-planet-simulations}

Ideally, we aim to test how the detection methods perform over the
possible range of variation of all parameters.  However, the high
dimensionality of the problem of determining astrometric and Keplerian
parameters together prevents a full exploration of all parameters on a
grid.

For our single-planet simulations, we selected three parameters to be explored on a grid:
\begin{itemize}
\item the signal-to-noise ratio S/N;
\item the period $P$;
\item the eccentricity $e$.
\end{itemize}
All the other parameters were randomised using the probability
distributions listed in Table~\ref{table:probdistr-astropar} and
explained in Sect.~\ref{sec:astrometricparms}. 

We produced a first batch of simulations for a 5 yr mission (the
nominal Gaia mission length), and a second batch for a 10 yr mission
to investigate the improvements in case the mission is extended. In
both batches, our grids ran with logarithmic increments over the
ranges $\SN=0.2$--10, $P=0.08$--10 yr, and $e=0.01$--0.97. To smooth the
grid and make the plots more readable, we added small random amounts
to the S/N, $P$, and $e$ of each simulation. The upper bound to the
eccentricity corresponds to an ellipse with a 1:4 ratio between the
semi-minor and semi-major axis.

\subsection{Randomised astrometric parameters}
\label{sec:astrometricparms}

\begin{table}
  \caption[]{Probability distributions of randomised parameters for
    the single-planet simulations. The parameters for the
      normal distribution are mean and sigma; for the log-normal they
      are log-mean and sigma; for the uniform they are the bounds of
      the allowed interval.}
  \label{table:probdistr-astropar}
  \centering                       
  \begin{tabular}{ll}
    \hline\hline      
    Parameter      &  Distribution \\
    \hline                       
    RA, Dec                      &uniform over the celestial sphere \smallskip \\
    $\Delta \alpha_0$ (mas)      &normal(0, 0.04) \\
    $\Delta \delta_0$ (mas)      &normal(0, 0.04)\\
    \multirow{2}{*}{$d$  (pc)}   &log-normal($\ln \frac{1\,
                                          \mathrm{mas}}{\sal\ \SN}$, 0.6) \\
    $\varpi$  (mas)              &1000/d \\
    \multirow{2}{*}{$\mu_{\alpha}$, $\mu_\delta$ (mas yr$^{-1}$)}  &taken randomly from Gaia~DR1 \\
                                 &stars with distance $\sim d$ \\
    $T_p$     (yr)         &uniform(0, $P$)  \\
    $\Omega$           &uniform(0, $180^\circ$)   \\
    $\omega$           &uniform($-180^\circ$, $180^\circ$)   \\
    $\cos i$          &uniform($-1$, 1)         \\
    \hline                       
  \end{tabular}
\end{table}

We set the RA and Dec coordinates of the host star to random values,
uniformly distributed over the celestial sphere. The Gaia scanning law
was computed for each position using the AGISLab software
\citep{holl2012b-agislab}, which was developed as part of the Gaia
astrometric global iterative solution (AGIS, \citealt{lindegren2012})
and is available to users within the Gaia Data Processing and Analysis
Consortium (DPAC).

We set the astrometric parameters ($\Delta \alpha$, $\Delta \delta$,
$\varpi$, $\mu_{\alpha}$, $\mu_\delta$) to random values according to
the probability distributions listed in
Table~\ref{table:probdistr-astropar}. We assumed $\Delta \alpha$ and
$\Delta \delta$ to be normally distributed, with a standard deviation
of 1 mas. The remaining parameters depend on the distance $d$, which
was assumed to be log-normally distributed around a value which
depends on the S/N. The range of values covered by $d$ was chosen so
that the S/N could include the cases of a Jupiter-mass planet at 3 AU
around a $1 \msun$ star, and of a Neptune-mass planet at 4 AU around a
$0.5 \msun$ star.  The parallax was obtained from the distance.  The
proper motions should reflect the larger spread in velocities of
nearby stars with respect to distant stars; therefore, we drew proper
motions from the Gaia data release 1
\citep[DR1;][]{gaia-dr1-summary,gaia-dr1-astrometry}.

We set the remaining orbital elements ($T_p$, $\Omega$, $\omega$, $i$)
to random values, different for each simulation, uniformly distributed
over their natural ranges of variation: $0\le T_p < P$, $0\le\Omega <180^\circ$,
$-180^\circ\le\omega <180^\circ$, $-1\le \cos i \le 1$.

Kepler's equation needs to be solved for every
simulated Gaia scan; we used the numeric solver \texttt{vpasolve} in
the MATLAB package, which provides solutions with double-precision accuracy%
\footnote{
We also experimented with the \citet{mikkola1987} cubic approximation,
which has a relative error within 0.2\%. We found that the use of the
approximation was hindering the successful fit of the simulations with
the method described in Sect.~\ref{sec:bayes}.
}.
\subsection{Simulations of systems with no planet}
\label{sec:no-planet-sim}

To help set up the thresholds for planet detection, we also simulated
no-planet systems. These systems follow the same specifications of the
previous sections for the astrometric parameters, noise, and
scanning law, but no Keplerian signal is added.

\section{Bayesian analysis}
\label{sec:bayes}

In this section, we assume some familiarity with Bayesian inference
and its nomenclature (prior, likelihood, posterior, Markov chains,
etc). The interested reader can find an introduction geared towards
astrophysics in \cite{trotta2008} or a more general discussion in the
textbook by \cite{gregory2005bayesian}.

\subsection{Modelling}
\label{sec:stanmodel}

\begin{table}
  \caption[]{Priors for the purely astrometric model. The parallax
    ($\varpi$) has the additional constraint that it must be
    non-negative.
    Proper motions have a parallax-dependent standard deviation.
    Distribution parameters as in Table~\ref{table:probdistr-astropar}.
  }
  \label{table:priors-astro}
  \centering                       
  \begin{tabular}{ll}
    \hline\hline      
    Parameter          & Distribution                          \\
    \hline                       
    $\Delta\alpha_0$ (mas) & normal(0, 0.04)    \\
    $\Delta\delta_0$ (mas) & normal(0, 0.04)                 \\
    $\varpi$         (mas)  & normal(2.8 , 59); $\varpi\ge 0$     \\
    $\mu_{\alpha}$   (mas yr$^{-1}$)& normal( 0, 9.5\,$\varpi^{0.89}$)            \\
    $\mu_{\delta}$    (mas yr$^{-1}$)  & normal( 0,10\,$\varpi^{0.83}$)            \\

    \hline
  \end{tabular}
\end{table}

\begin{table}
  \caption[]{Priors for the orbital elements of the planet. The full
    model includes both the orbital elements and the astrometric
    parameters listed in  Table~\ref{table:priors-astro}. The parameters for
    the von Mises distribution are location and concentration
    ($\kappa$). Other distributions are as in Table~\ref{table:probdistr-astropar}.
    The eccentricity is bound
    to the [0,0.9999] interval whose upper bound corresponds to an
    ellipse with a $1:70$ ratio between the semi-minor and semi-major axes.
  }
  \label{table:priors-full}
  \centering                       
  \begin{tabular}{ll}
    \hline\hline      
    Parameter            & Distribution                                    \\
    \hline              
    $P$     (yr)             & $\ln (P/\mathrm{1\ yr})$ is uniform($-6, 5$)  \\
    $\upsilon$ (mas)                 & log-normal($ -2, 1)$                      \\
    $\Omega$             & von Mises(0, 0.001, support=$[0, 180^\circ])$                                \\
    $\omega$             & von Mises(0, 0.001, $[-180^\circ, 180^\circ])$                             \\
    $\cos i$            & uniform($-1,1)$                                  \\
    $e$                  & uniform($0, 0.9999$)   \\
    $T_p$     (yr)           & von Mises(0, 0.001, $[0,P]$)         \\
    \hline 
  \end{tabular}
\end{table}

From a model point of view, Bayesian inference works similarly to
maximum likelihood in that a series of tentative sets of parameters
are generated, and each set is used to simulate an observable quantity
which is then compared against the data.  In our setting the data are
$\eta_{0,s}$ i.e.\ the observed AL displacements, one per Gaia scan $s$,
computed at simulation time using
Eq.~(\ref{eq:eta-astrometric-keplerian}). Let $\eta_{i,s}$ be the
displacements computed during the fit for a set $i$ of astrometric and
orbital parameters.  The likelihood can then be obtained by assuming
that the difference $\eta_{0,s} - \eta_{i,s}$ is normally distributed,
with mean zero and standard deviation $\sal$:
\begin{equation}
  \label{eq:likelihood}
  \mathcal{L}(i) = \frac{1}{\sqrt{2\pi}\sal} \prod_s 
     e^{-\frac{1}{2}\frac{(\eta_{0,s}-\eta_{i,s})^2}{\sal}} \quad .
\end{equation}

We used the \stan\ package for statistical inference \citep[][version
2.12]{carpenter2016stan} to model our simulated Gaia observations. The
\stan\ package allows models to be specified in a programming language
very similar to that of popular packages such as BUGS
\citep{lunn2000winbugs} or JAGS \citep{plummer2003jags}. All these
packages implement general-purpose MCMC samplers together with a
high-level programming language to specify the model.

Our implementation follows the formalism of Sect.~\ref{sec:orbits}.
For each simulated system, we define two models:
\begin{itemize}
\item a purely astrometric model, which only contains
  $\Delta\alpha_0$, $\Delta\delta$, $\varpi$, $\mu_{\alpha}$, and
  $\mu_\delta$. This model can be obtained by setting $A=B=F=G=0$ in
  Eq.~(\ref{eq:eta-astrometric-keplerian}). This model is the baseline
  against which the full model is compared when evaluating a
  detection;
\item a full model, containing all the parameters of
  Eq.~(\ref{eq:eta-astrometric-keplerian}).
\end{itemize}

Kepler's equation needs to be solved at each MCMC iteration, but
\texttt{Stan} does not provide a solver in its current version;
therefore, we use the \citet{markley1995} approximation, which is fast
to compute and provides, in our implementation, a relative error of
the order of $10^{-14}$, close to the limits of a double-precision
variable.

\subsection{Model priors}
\label{sec:priors}

\subsubsection*{Purely astrometric model}

The prior probability distributions for the model parameters should be
set according to the expected range of variation, and to the expected
distribution for their values.

For the astrometric model we chose very weakly informative priors
which only set the physical scale of the parameters
(Table~\ref{table:priors-astro}).  For $\Delta\alpha_0$ and
$\Delta\delta$ we used normal distributions with {mean 0 and
standard deviation 0.04 mas, consistently with the expected
perturbation induced by a Jupiter-mass planet. For the proper motions
$\mu_{\alpha}$ and $\mu_\delta$ we used normal distributions with
mean 0 and standard deviation that depends on the
parallax to reproduce the fact that more distant stars have lower
velocities. By fitting the Gaia~DR1 stars within $\sim 200$~pc, we
found $\sigma_{\alpha_0}\propto \varpi^{0.83}$ and
$\sigma_\delta\propto \varpi^{0.89}$.  For the parallax $\varpi$, we
used a log-normal prior (with ln(mean)$=2.8$, corresponding to
16 mas, and ln(standard deviation)$=7$) which we found to
reasonably reproduce the Hipparcos and Gaia~DR1 data. We also added
the constraint that $\varpi\ge 0$.

\subsubsection*{Full model}

For some parameters ($\Omega$, $\omega$, $\cos i$, and $T_p$) the
obvious choice is to set the priors to constant distributions
according to the parameter range (see Table~\ref{table:priors-full}).
A uniform distribution is therefore our choice for $\cos i$.  However,
$\Omega$ and $\omega$ are angles, and $T_p$ can also be treated as
such, thus one needs to ensure that the prior distribution is defined
over a circular support so as not to introduce any artificial barrier for
when the Markov chains explore the likelihood space. This is achieved
by a von~Mises distribution, with concentration parameter
$\kappa\sim 0$, which is the circular equivalent of a uniform
distribution\footnote{The von Mises distribution is a circular
    analogue of a normal distribution. It is a continuous probability
    distribution whose support is a circle, and it is defined as
    $
      \mathrm{von\ Mises}(x; \mu_\mathrm{vM},\kappa) = 
      \frac{e^{\kappa \cos(x - \mu_\mathrm{vM})}}{2\pi I_0(\kappa)},
    $
    where $I_0(\kappa)$ is the modified Bessel function of order 0.
    The parameters are the location of its peak $\mu_\mathrm{vM}$ and
    the concentration $\kappa$ (analogous to the inverse of the
    variance of a normal; distributions with higher $\kappa$ have
    sharper peaks). When $\kappa=0$, no peak is present and the
    location becomes undefined. Ideally, one should set $\kappa=0$;
  however, the current Stan implementation of the von~Mises
  distribution is restricted to $\kappa>0$, so we chose a $\kappa$ low
  enough to be practically indistinguishable from 0.}
\citep{jammalamadaka2001}.

For the other orbital parameters we chose non-informative priors as follows:
\begin{itemize}
\item For $P$, we considered a prior where $\ln P$ is uniformly
  distributed, so that no preference for any scale is
  introduced. Periods shorter than 1 day or longer than 150 yr were
  excluded.
\item For $e$, we also considered a uniform prior. An upper bound
  $e=1$ cannot be used because it corresponds to an open
  orbit. Therefore, we took $e=0.9999$ as the upper bound (which
  corresponds to an ellipse with a ratio between the major and minor
  axes of $\sim 70$).
    
\item For the angular size of the semi-major axis $\upsilon$ we chose
  a log-normal distribution with ln(mean)$=-2$ (corresponding to 0.14
  mas) and ln(standard deviation)$=1$, with an upper bound at 10
  mas. This distribution puts 90\% of the probability on systems
  with $\upsilon<0.5$ mas, yet it allows for some rarer systems closer
  to Earth where wider angular orbits might be expected.

\end{itemize}

Finally, astrometric parameters in the full model have the same priors as in
the purely astrometric model.

\subsection{Chain initialisation}
\label{sec:chi2}

Markov chains (MC) need a starting point. From there, the MC will
start to explore the parameter space. If the starting point is not in
a high-probability region, it is necessary to discard the first
iterations (usually called the burn-in period or warm-up period) to
ensure that the rest of the chain represents a sample of the posterior
distribution, with no bias due to the presence of many initial steps
exploring a low-probability region.

In principle, the starting point can be taken randomly, though
this may result in a very long burn-in period. Therefore, statistical
literature often suggests  starting the Markov chains at a point
reasonably close to the likelihood global maximum if it is known
\citep[e.g.][]{geyer2011mcmc}.

Our tests have shown that the MC burn-in can be prohibitively long for
the needs of a survey unless a starting point close to the correct
values for the parameters is used. However, in our case the structure
of the likelihood space presents many very narrow
peaks\footnote{The highest secondary peaks are randomly located and do not
    seem to be aliases of the true period; there  are usually only a few
    for very high S/N (e.g. S/N$\sim 10$). In addition, a large number
    of lower peaks are present even at S/N$\sim 0$, which can just be
    attributed to noise. The peak narrowness does not depend on the
    S/N.}, which prevent common optimisation methods (e.g.\ Amoeba,
Levenberg-Marquardt, or Broyden–Fletcher–Goldfarb–Shanno) from finding
the global likelihood maximum.

We have considered two possibilities for the starting point: first, a
perfect starting point corresponding to the true value of the
parameters; second, a best-fit point obtained from a simplified model
that assumes circular orbits. The first option is motivated by our aim
to test our Bayesian model and check how it behaves in an ideal
case. The second option is more realistic; our motivation is to test
whether a very simple and fast model can be used as a reasonable
starting point or if a more refined approximation is needed. In the
reality of any future survey using Gaia data, the first option will of
course be unavailable, while more refined methods used to find a starting
point can be used \citep[e.g.][]{sahlmann2013}.

In the circular orbits model, the period
$P$ is the only parameter which appears non-linearly. We defined a
grid of frequencies $f=1/P$, starting from $f=0.067$
and increasing to 12.5 yr$^{-1}$,
with a step of 0.03 yr$^{-1}$, covering the allowed interval for
$P$. For every value of $f$, we obtained the best-fit values for all
other parameters and the $\chi^2$ using linear least-squares.  The
values which scored the lowest $\chi^2$ provided the starting point
for the MC.

The circular-orbit model only includes ten parameters: $e$
does not appear, and $\omega$ and $T_p$ become degenerate, so that only
one of them is needed. To make our implementation easier, we dropped
$T_p$ and retained $\omega$. Therefore, we were left with the need to
specify starting values for $e$ and $T_p$. We chose $e=0.1$ and $T_p=0$.

\subsection{Chain convergence}
\label{sec:chain-convergence}

In the long run, the set of points touched by a MC forms a non-biased
sample of the target probability distribution; when this happens, it
is said that the MC has converged to the target.  The main risk
associated with non-converging chains is that, since the MCMC samples
would not be representative of the posterior distribution, any
inference on the model parameters might be severely biased.  The
purpose of the burn-in period is to assure that the chains have
converged before  posterior samples are obtained  from
them. However, there is little guidance on how long the burn-in period
should be, so one has to rely on experience and on inspection of the
MCMC samples.

We initially conduced some test runs of our \texttt{Stan}
implementation to understand the number of iterations needed for the
Markov chains burn-in period. Visual inspection of the chain traces
showed that the burn-in period lasted less than a few hundred
iterations in many cases where $\SN\gtrsim$3--5, but in some cases it could
be considerably longer. 

Therefore, for the simulations described in
Sect.~\ref{sec:single-planet-simulations}, we initially decided to use
a burn-in period of 500 iterations.  Then we  checked whether the chain
had converged after this period, or if more burn-in iterations were
needed, according to the procedure described as follows.

After the 500 burn-in iterations, we collected the next 500 iterations
for further analysis. On the latter samples we computed for each
parameter $i$ the potential scale reduction $\hat R_i$, which measures
how much sharper an estimate might become if the MCs were run
indefinitely (\citealt{gelmanrubin1992}; Eq.(11.4) in
\citealt{BDA3}). The $\hat R$ compares the variance among the chains
versus  the variance within a single chain: if all chains have converged
and are sampling the same space, both variances should be
approximately equal, and $\hat R\sim 1$.  A value of
$\hat R\gtrsim 1.1$ is usually considered as pointing towards
convergence problems. The opposite is not necessarily true (it is possible
  to have $\hat R< 1.1$ and still have convergence problems) because
  convergence can never be guaranteed when the number of samples is
  finite, though in our experience the $\hat R$ criterion has matched very
  well any conclusion that could be made based on visual inspection.

We found that, for any given simulated system, the $\hat R_i$ can vary
among the parameters, with some of them showing convergence problems,
while other seem to be fine. Therefore, we also computed the
average $\hat R_\mathrm{avg} = \frac{1}{N} \sum_{i=1}^N \hat R_i$.  Whenever a
simulated system was found to have $\hat R_\mathrm{avg} > 1.1$, we
discarded its samples and we ran the chains anew, this time with 15000
burn-in iterations.

After that, if $\hat R_\mathrm{avg}$ was still unsatisfying, we deemed
the chains as not converging and excluded the planet from further
consideration.

There may be different reasons for not fulfilling the convergence
criterion, some of which do not prevent meaningful information to be
retrieved. However, it would be  difficult to express how confident one
could be in any estimate coming from non-converging chains. Therefore,
in the context of a blind survey of exoplanets, information from
non-converging chains might be of little use;  in the context of
extracting the most information from Gaia data of a particular planet
(say, by joint fitting of astrometric and radial velocity, or transit
data) there would  probably be more prior information to help find a
different (perhaps better) starting point for the chain.

\subsection{Exoplanet detection as model selection}
\label{sec:detection}

Deciding whether an exoplanet has been detected is essentially a
comparison of model performances. The null model is that no planet is
present, and that the Gaia measurements can be adequately explained
with just the astrometric parameters. The alternative model is that a
planet is present, and both the astrometric and Keplerian parameters
are needed.

The classical (frequentist) approach is to use hypothesis testing: the
difference between the models' $\chi^2$ is computed and linked,
through the number of degrees of freedom, to a p-value. It can be
applied to the case of exoplanet detection, provided that the
Thiele--Innes parameterisation is adopted.
 One would therefore obtain the $A$, $B$, $F$, and $G$ parameters from
the fit, and only later would they  be reconnected to the orbital
elements $\upsilon$, $\omega$, $\Omega$, and $\cos i$. Uncertainties on
the best-fit values should then be propagated from the
Thiele--Innes parameters back to the orbital elements.

However, a complication is that the number of degrees of freedom is
unambiguously defined only for linear models. Non-linear models behave
as if they have a larger number of degrees of freedom than a linear
model with the same number of parameters. Also, the `effective degrees
of freedom' of a non-linear model are data-dependent (through the
variance of the parameters; see e.g.\ Eq.~(\ref{eq:waic}) below), and
they are larger when the data are noisy \citep{andrae2010}; we will
show this happening in our simulations in Sect.~\ref{sec:dchisq}.

Conversely, the Bayesian approach has the following advantages: the
orbital elements can be the subject of inference, the uncertainties on
the fit parameters need not  be estimated using asymptotic theory,
and there are model comparison methods which explicitly account for
the effective degrees of freedom.

An easy way to perform Bayesian model comparison is to use information
criteria (IC) because only the posterior samples are needed for their
calculation. The theory of most IC is based on the notion of
predictive accuracy, i.e.\ how well a model fit can predict new data
produced from the true data-generating process. A notable exception is
the Bayesian information criterion (BIC, see below) which is motivated
by the marginal probability of the data under the model. A
introduction to both criteria from an astrophysics perspective can be
found in \citet{liddle2007}; among the statistical sources, we refer to
\citet[Chap.~7]{BDA3} and \citet{burnham-anderson}.

The IC operate in practice by computing a function of the likelihood,
either at a specified point or averaged over the posterior.  The
function includes a penalty, based on the number of degrees of
freedom, to account for overfitting by the more complex model.

Some IC also find application in the context of frequentist
statistics as they can be applied to maximum likelihood
estimates. Therefore some of our results might also find application 
in future work that does not use the Bayesian framework.

The best known IC are the Akaike information criterion
\citep[AIC,][]{akaike1973},
\begin{equation}
  \label{eq:AIC}
  \mathrm{AIC} = -2 \ln \mathcal{L}_\mathrm{max} + 2 k
,\end{equation}
and the Bayesian information criterion \citep[BIC,][]{schwarz1978bic},
\begin{equation}
  \label{eq:BIC}
  \mathrm{BIC} = -2 \ln \mathcal{L}_\mathrm{max} + k \ln n
,\end{equation}
where $\mathcal{L}_\mathrm{max}$ is the maximum likelihood, $k$ is the
number of free parameters, and $n$ is the number of data points. The
main difference is that the penalty on BIC also depends on the number
of data points.

Both the AIC and the BIC rely on a point estimate of the model, and
they still involve counting the free parameters. A recent development
on the AIC is the widely applicable (or Watanabe--Akaike) information criterion
(WAIC), which has been proposed as a fully Bayesian approach of
estimating the predictive accuracy (\citealt{watanabe2010}; see
\citealt{ranalli2016} for an application to astrophysics). It uses all
available samples of the posterior distribution and gets the effective
degrees of freedom from the parameter variance:
\begin{align}
  \label{eq:waic}
  \begin{split}
  \mathrm{WAIC} &=
    -2 \ln \left( \frac{1}{I} \sum_{i=1}^I \mathcal{L}(i) \right)  \\
    &+2 \sum_{s=1}^n \frac{1}{I-1} \sum_{i=1}^I 
      \left(
        \ln \mathcal{L}(i,s) - \langle\ln \mathcal{L}(i,s)\rangle_i
      \right)^2.
  \end{split}
\end{align}
Here,
the index $i$ runs on the samples from the posterior distribution, and
the index $s$ runs on the data points (the Gaia scans), so both
indices keep the same meaning they had in
Eq.~(\ref{eq:likelihood}). The symbol $\mathcal{L}(i,s)$ is the
likelihood of a single data point (scan). The angular brackets with the
$i$ subscript indicate that the average over the samples should be
taken; the squared difference is therefore the variance of the
log-likelihood%
\footnote{\citet{watanabe2013wbic} also proposed an
  extension of the BIC,  the widely applicable Bayesian information
  criterion (WBIC).  However, to compute the WBIC  samples
  from both the posterior and the prior are needed. The latter are not provided
  by the current Stan version, and therefore we do not consider the
  WBIC here.}.

In all cases, the lower the IC, the more favoured the model. However,
the absolute scale of the IC does not carry statistical meaning; the
important quantity is the difference between the IC of different
models, not the values of the IC themselves. Therefore, a large $\Dic$
can indicate a strong preference for one model over the other, while a
small $\Dic$ indicates that one model does not perform appreciably
better than the other. 

The $\Dic$ can also be interpreted using Jeffreys' scale of evidence
\citep{efron2001}, which rates $\Dic\gtrsim 1$, 3, 20, and 150 as
`barely worth reporting', `weak', `strong', and `very strong',
respectively.

The  preference can be quantified by considering the relative
likelihood $w$ of model $M_1$ over model $M_2$. Assuming
IC$(M_1)>$IC$(M_2)$ (so that $M_2$ is the preferred model),
\begin{equation}
  \label{eq:model-likelihood}
  w = e^{-\frac{1}{2}\, \Dic_{12}}
,\end{equation}
where $\Dic_{12}=\mathrm{IC}(M_1) - \mathrm{IC}(M_2)$ \citep[see their
Sects.~2.8 and 2.9]{burnham-anderson}. Strong evidence, in the
  sense of Jeffreys' scale, would correspond to a relative
likelihood of $w\sim 4.5\e{-5}$ for the worse model over the preferred
one.

Reflecting the different definitions of the IC, the actual values of
$w$ will depend on which IC is considered. The most important factor
is the $\ln\ n$ term that appears in BIC (Eq.~\ref{eq:BIC}), but not in
AIC or WAIC (Eqs.~\ref{eq:AIC},\ref{eq:waic}). The BIC in fact has the
property that in the limit of large $n$ the penalty tends to
infinity. The AIC and WAIC do not have this property.

The importance of this property is more evident if one considers
the case of a real multiplanet system whose data are fit by a set of
models that approximate our physics knowledge better and better (e.g.
no planet; one planet; two planets ignoring their mutual attraction;
two planets including their mutual attraction; three planets
\dots). Obviously, some models are more easily distinguished (no
planet vs.\ one planet), while some others are more difficult to
distinguish (mutual attractions).  In the limit of large $n$, BIC will
discard models that overfit data more easily than AIC or WAIC, while
AIC and WAIC will instead identify a subset of similarly behaving
models. Whether this property of BIC is desirable depends on whether one
expects that the `true' model belongs in the considered set. One
seldom or probably never has such an expectation in
general. However, planetary motions may represent a niche case where
one can expect to include a model that, for a given data quality, is
indistinguishable from one that includes all known physical effects.

For Gaia data, $n=73\pm23$ and $147\pm 46$ for the 5~yr and 10~yr
mission, respectively, which means that from Eqs.~(\ref{eq:AIC}) and (\ref{eq:BIC})
one expects an average $\Delta \mathrm{AIC}-\Delta\mathrm{BIC}\sim 16$
and 21, respectively. These values are close to the threshold for
strong evidence on Jeffreys' scale. Therefore, we expect that different
IC may yield different fractions of false positives if the strong
evidence threshold is used. From the defining equations of the IC and
for systems with S/N$\sim 0$, we expect that the BIC signals a
preference for the simpler model (the astrometric model), while the
AIC and WAIC signal that the more complex model (the planetary model)
does not offer an advantage over the simpler model.

One might ask what advantage the IC offer over a $\dchisq$
criterion. In a non-linear setting, because of the difficulties in
determining the number of degrees of freedom, a $\dchisq$ is
equivalent to an information criterion where one does not
know beforehand what penalty to apply to correct for overfitting. It
would still be possible to use the $\dchisq$ for model choice after
calibrating a threshold based on an allowed fraction of false
positives.

\section{Fraction of false-positive detections}
\label{sec:bayes-results-1pl}
\label{sec:falsepositives}

\begin{figure*}
  \centering
  \includegraphics[width=.9\columnwidth]{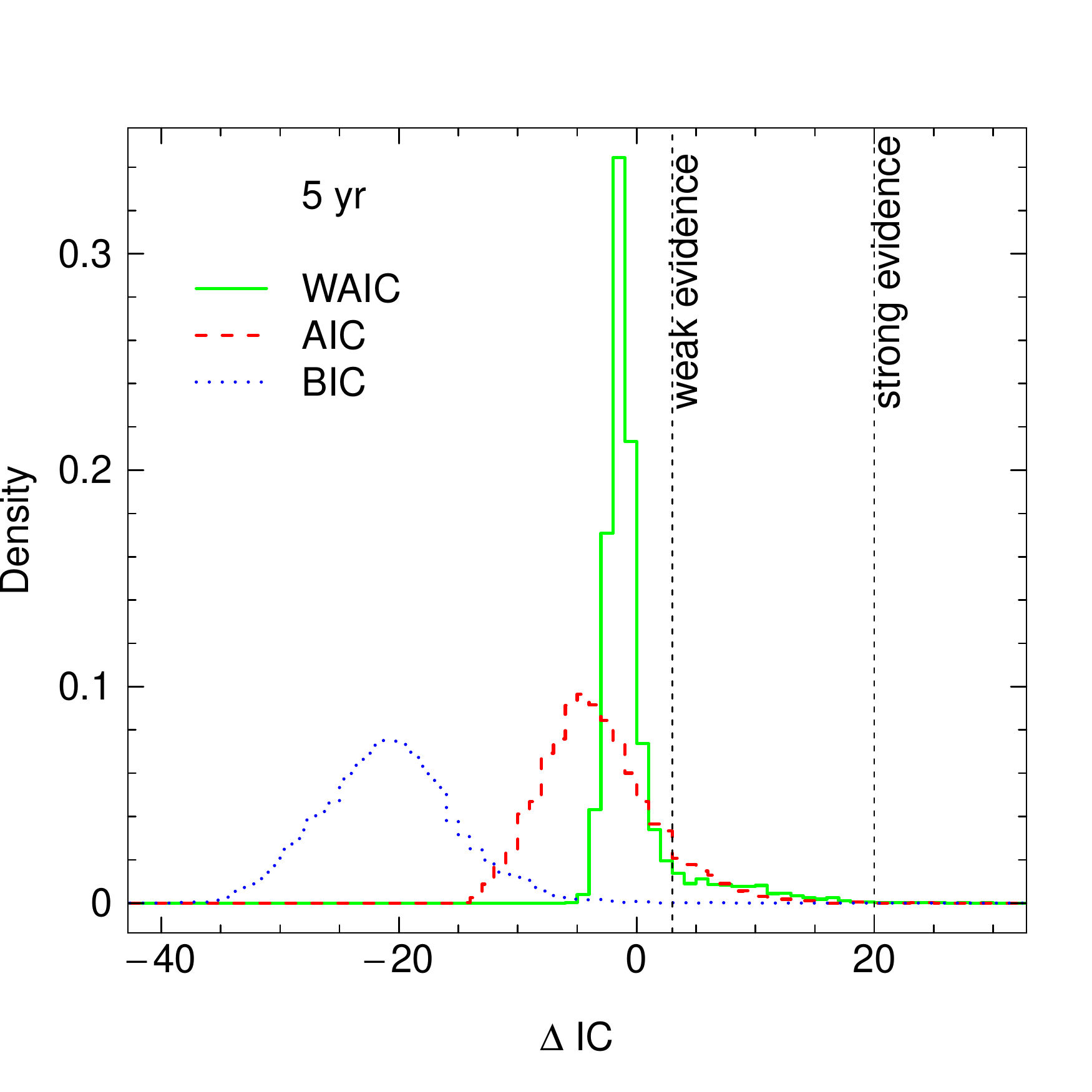}
  \includegraphics[width=.9\columnwidth]{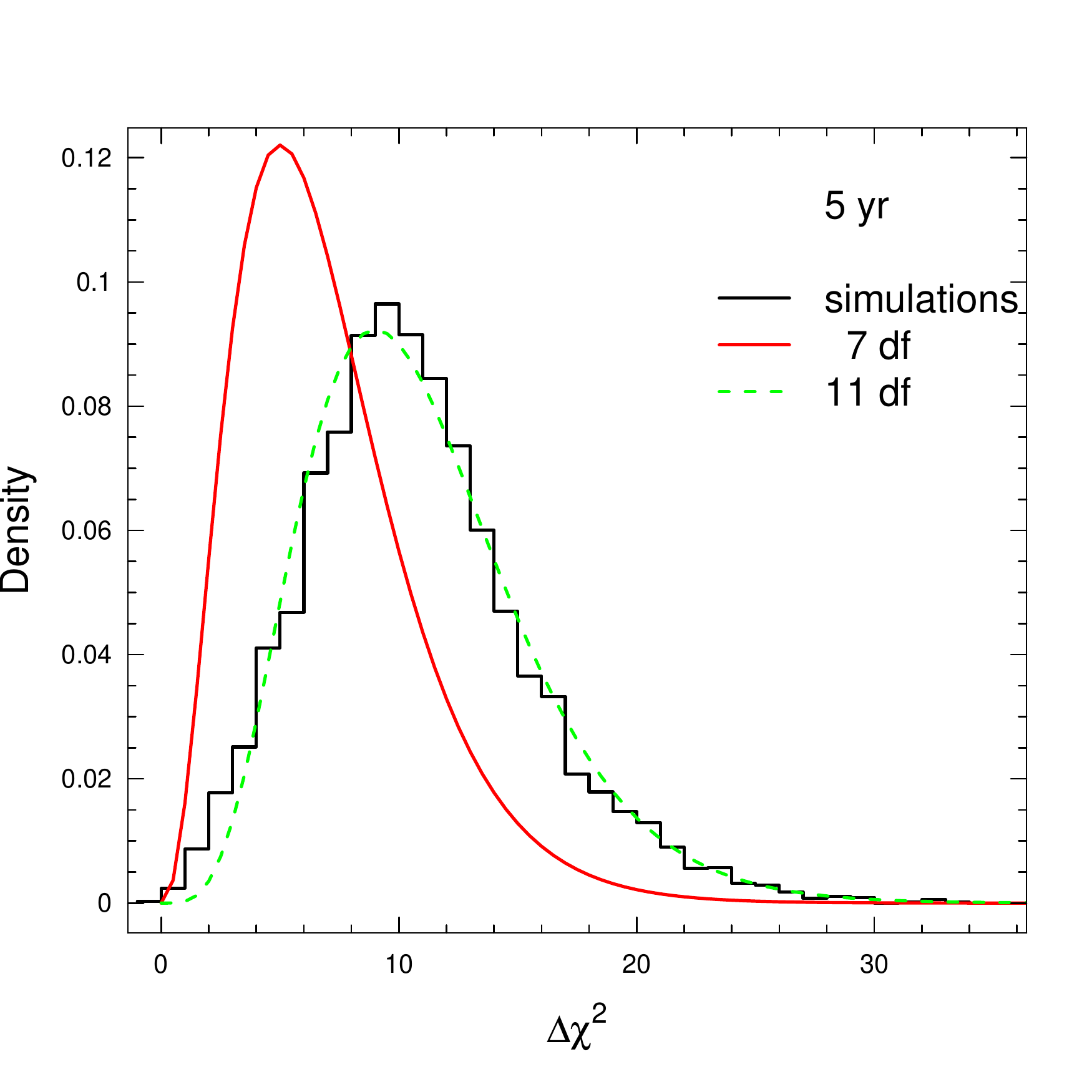}\\
  \includegraphics[width=.9\columnwidth]{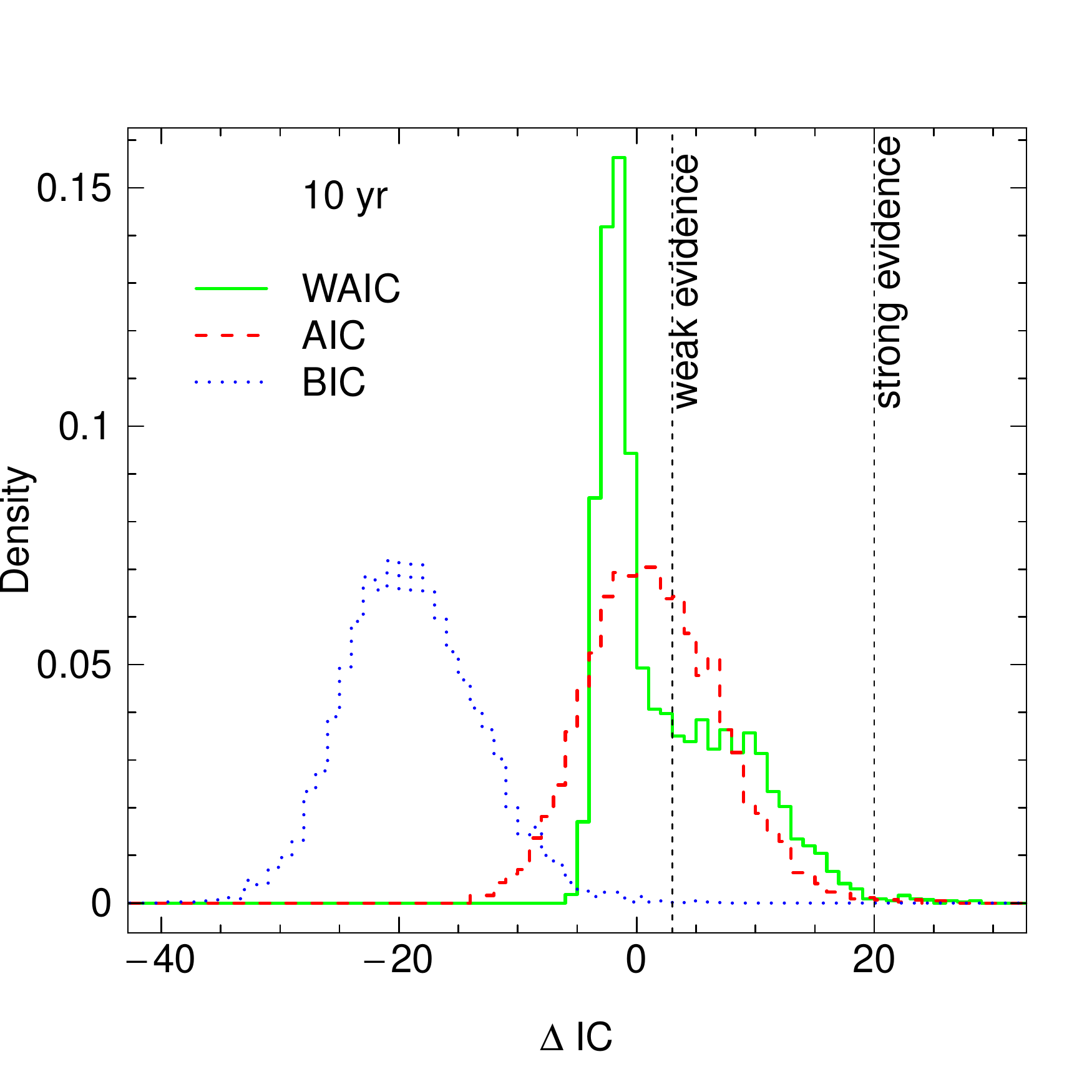}
  \includegraphics[width=.9\columnwidth]{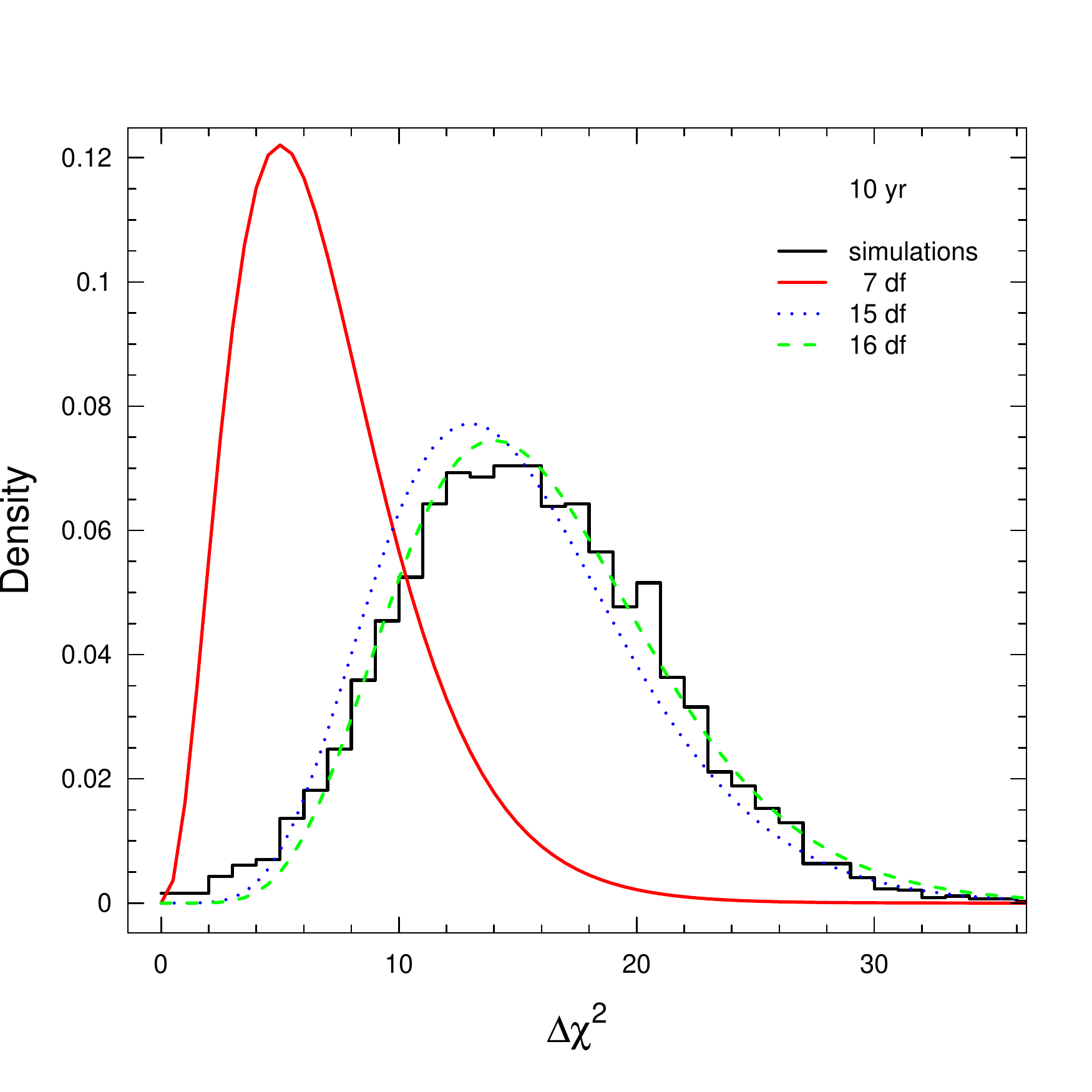}
  \caption{Histograms for the  IC (left panels) and for $\dchisq$
    (right panels) in the case of no-planet simulations. The upper
    panels consider the 5~yr mission, the lower panels the 10~yr
    mission.
    Left panels: Green solid lines, WAIC; red dashed lines, AIC; blue
    dotted lines, BIC;  the
    vertical dashed lines show the IC thresholds for weak and strong
    evidence according to Jeffreys' scale.
    Right panels:
    Solid and dashed curves show
    $\chi^2$ distributions with the degrees of freedom
    (dof) reported in the legend. The empirical density does not follow a
    $\chi^2$ distribution with the nominal number of dof (7, solid red line). It
    can however be approximated, by a
    $\chi^2$ distribution with a larger number of dof: 11~dof for the
    5~yr mission, 15~dof (or 16~dof for $\dchisq\gtrsim 30$) for the 10~yr mission.
  }
  \label{fig:ic_zerosn}

\end{figure*}

\begin{figure*}
  \includegraphics[width=.7\textwidth]{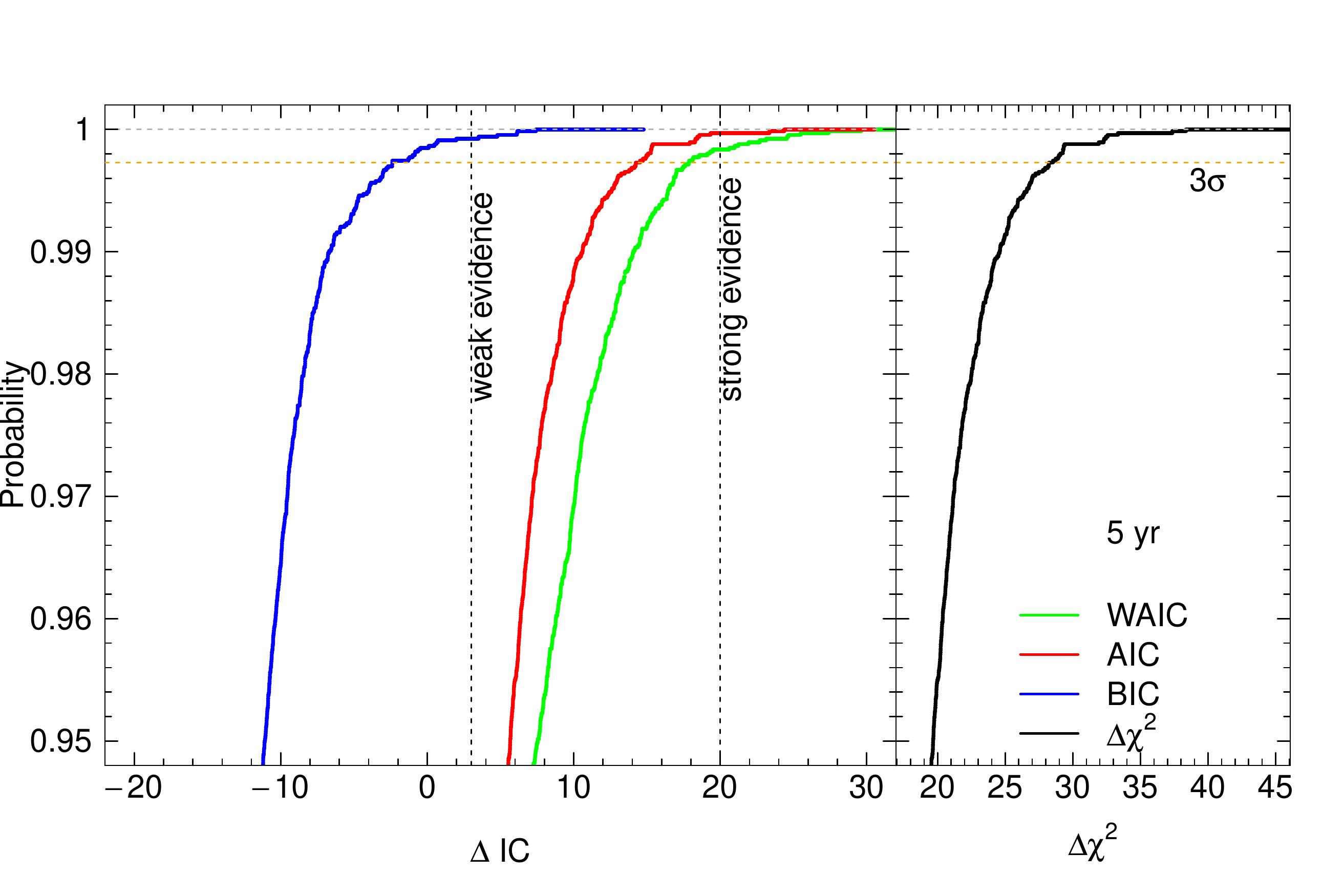}\\
 \sidecaption
  \includegraphics[width=.7\textwidth]{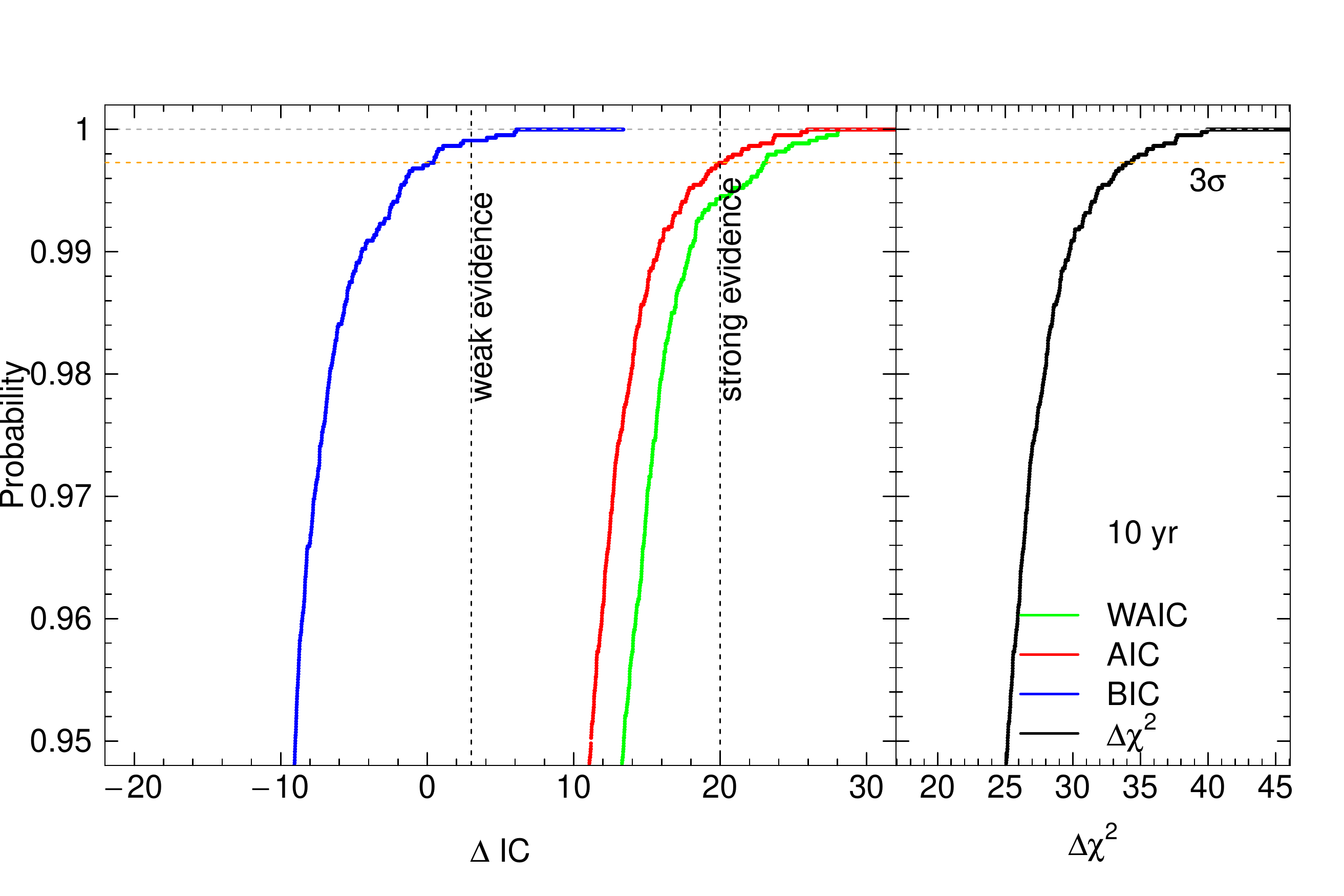}
  \caption{ Cumulative distributions of the IC and  $\dchisq$ for no-planet systems
    in the 5~yr (upper panel) and 10~yr (lower panel) mission
    simulations. The four curves show (from left to right): BIC (blue),
    AIC (red), WAIC (green), $\dchisq$ (black). The horizontal dashed
    lines show the 100\% and 99.73\% ($3\sigma$) quantiles. The
    vertical dashed lines show the IC thresholds for weak and strong
    evidence according to Jeffreys' scale.
  }
  \label{fig:cumunulldistr}
\end{figure*}

In this section we apply all criteria to our no-planet simulations, so
we expect that the purely astrometric model should be preferred in all
cases. Our samples consist of 6644 and 4402 no-planet systems for the
5~yr and 10~yr missions, respectively.

\subsection{Information criteria}
\label{sec:detection-thresh-ic}

\begin{table}
  \caption[]{Information criteria and $\dchisq$ quantiles for
      false-positive detections in no-planet simulations,
      for the four different indicators and for the
      5~yr and 10~yr missions. The first two columns list the 99\% quantile,
      while the last two columns show the 99.73\% ($3\sigma$) quantile
      values observed. The 99.73\% quantile has been estimated on a
      small number of objects so its value should be considered
      approximate.}
  \label{table:ICnulldistr}
  \centering                       
  \begin{tabular}{crrrr}
    \hline\hline
    Criterion & \multicolumn{2}{c}{99\% quantile} & \multicolumn{2}{c}{99.73\% quantile}\\ 
              & (5 yr)         &(10 yr)                         & (5 yr)                                  & (10 yr)\\
    \hline                                                                
    WAIC      & 14.0           &17.9                         & 18                                    & 23  \\
    AIC       & 10.6           &15.7                         & 14                                    & 20  \\
    BIC       & -6.72          &-4.51                         & -2.4                                   & 0.11\\
    $\dchisq$ & 24.6           &29.7                         & 28                                    & 34  \\
    \hline                       
  \end{tabular}
\end{table}

We define $\Dic=\mathrm{IC}(M_a)-\mathrm{IC}(M_k)$, where $M_a$ is the
astrometric model, and $M_k$ is the model with the Keplerian
orbit. With our definition, a positive $\Dic$ indicates a preference
for the Keplerian model, and vice versa. Values of $\Dic\sim 0$
indicate no preference for one model over the other; therefore, for
reasons of simplicity, the simpler model (i.e.\ the astrometric model)
should be preferred. For $\Dic>0$, we  use Jeffreys' scale of
evidence to interpret the results, and then we  define empirical
$\Dic$ thresholds based on the number of false positives.

In Fig.~\ref{fig:ic_zerosn} (left panels), we show the histograms
of $\Dic$ for the no-planet systems. The peaks of the histograms are
safely below the IC$>3$ threshold for weak evidence according to
Jeffreys' scale, though both AIC and WAIC have a tail in the
weak-evidence zone, especially in the 10~yr mission case. 
Such a tail may be better seen in the plots of the cumulative
distributions (Fig.~\ref{fig:cumunulldistr}).
For the 5~yr mission, the strong
evidence threshold (IC$>20$) is exceeded by AIC in 2 cases (0.03\%) and by WAIC
in 11 cases (0.17\%)
out of 6644 cases; for the 10~yr mission it is exceeded by AIC in 12
cases (0.27\%) and by WAIC in 25 cases (0.57\%) out of 4402. The
strong evidence threshold is never exceeded by BIC.
We list the 99\% and 99.73\%
quantiles in Table~\ref{table:ICnulldistr}. The value for the 99.73\%
quantile should be considered as approximate because of the small
number of systems whose ICs fall above the quantiles (18 and 12 for
the 5~yr and 10~yr mission, respectively).
When going
from the 5~yr to the 10~yr mission, there is an increase in
both quantiles for all ICs and for $\dchisq$.

Compared to $\Delta$AIC and $\Delta$BIC, $\Delta$BIC appears to be
negative in almost all cases and therefore indicates a preference for
the astrometric model.

\subsection{$\dchisq$}
\label{sec:dchisq}

Considering two nested linear models with a different number of
parameters (e.g. the astrometric and the full model), and
data distributed according to the null model (in our case, the
astrometric model), the difference between the $\chi^2$ values of the
best fits under the two  models still follows a $\chi^2$
distribution, with a number of degrees of freedom (dof) equal to the
difference between the dof of the two models. As discussed in
Sect.~\ref{sec:detection}, the dof become ill-defined when the model is
non-linear. We show this in Fig.~\ref{fig:ic_zerosn} (right panels),
where we plot the empirical $\dchisq$ density for the no-planet
simulations against a $\chi^2$ distribution for seven dof (corresponding
to the difference between 12 and 5 parameters in the full and
astrometric models, respectively). The empirical density is better
described by a $\chi^2$ distribution with 11~dof for the 5~yr mission;
and with 16~dof for the 10~yr mission (though 15~dof seem to better
reproduce the tail at $\dchisq\gtrsim 30$).

The cumulative distribution is shown in
Fig.~\ref{fig:cumunulldistr}. The 99\% and 99.73\% quantiles obtained
from the empirical distribution are listed in
Table~\ref{table:ICnulldistr}. As in the case of $\Delta$AIC and
$\Delta$BIC, the quantile increases when going from the 5~yr to the
10~yr mission.

\section{Detection rate}
\label{sec:detectionrate}
\label{sec:detect-thresh}

In this and in the following sections, we consider simulations of
one-planet systems. Our samples consist of 4968 and 4706 systems for the
5~yr and 10~yr missions, respectively.

We chose to put our detection threshold at $\Delta$BIC$=20$, which is
equivalent to $\Delta$AIC=70, $\Delta$WAIC$=70$, and $\dchisq=80$.
This has the effect that, with respect to $\Delta$AIC$=20$, the number
of false positives is minimised, and the quality of detection is
improved (our tests showed that in systems with very low S/N and hence
a contrasting AIC versus\ BIC verdict, the period was incorrectly
recovered in approximately one-quarter of the cases).

\begin{figure*}
  \centering

  \includegraphics[width=.9\columnwidth]{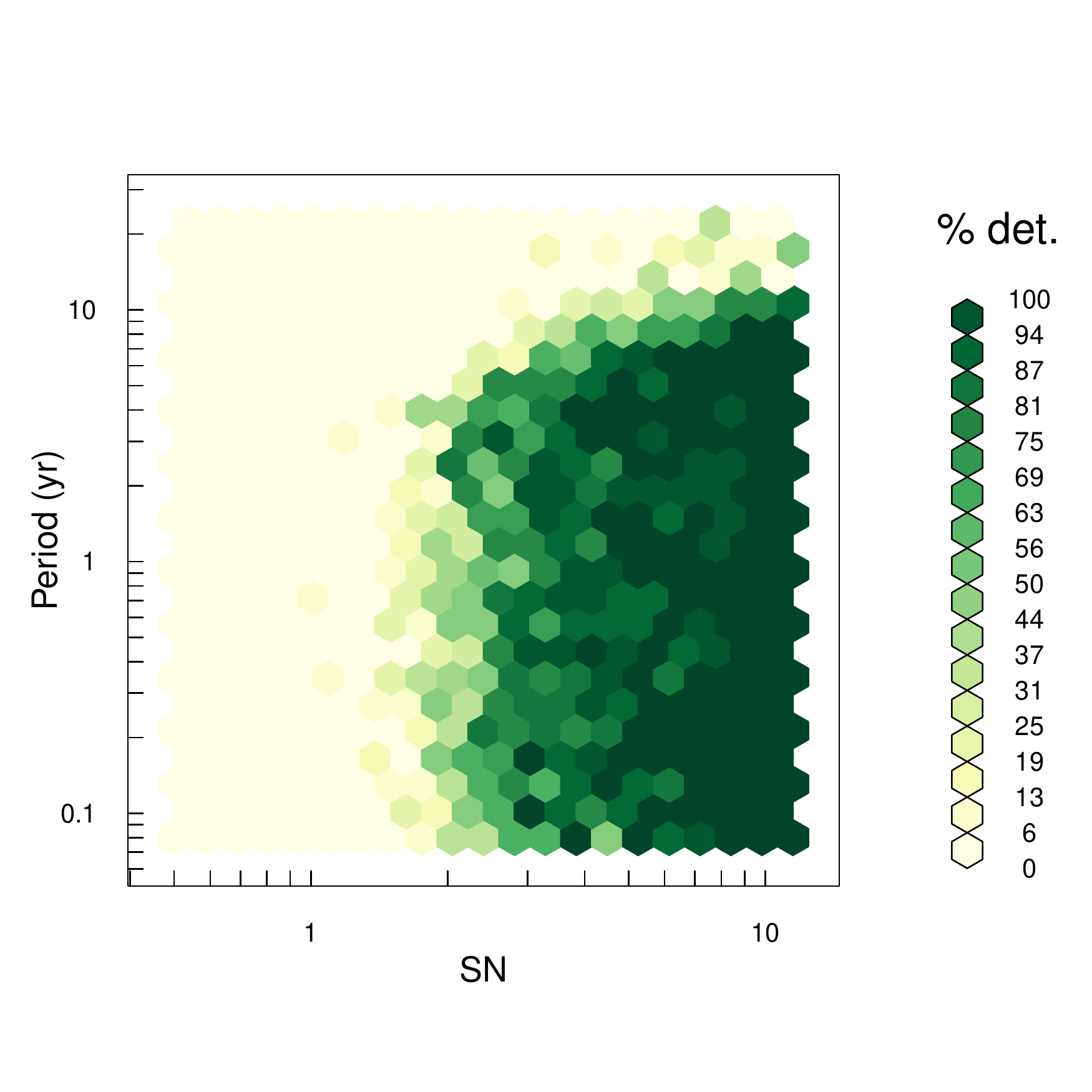}
  \includegraphics[width=.9\columnwidth]{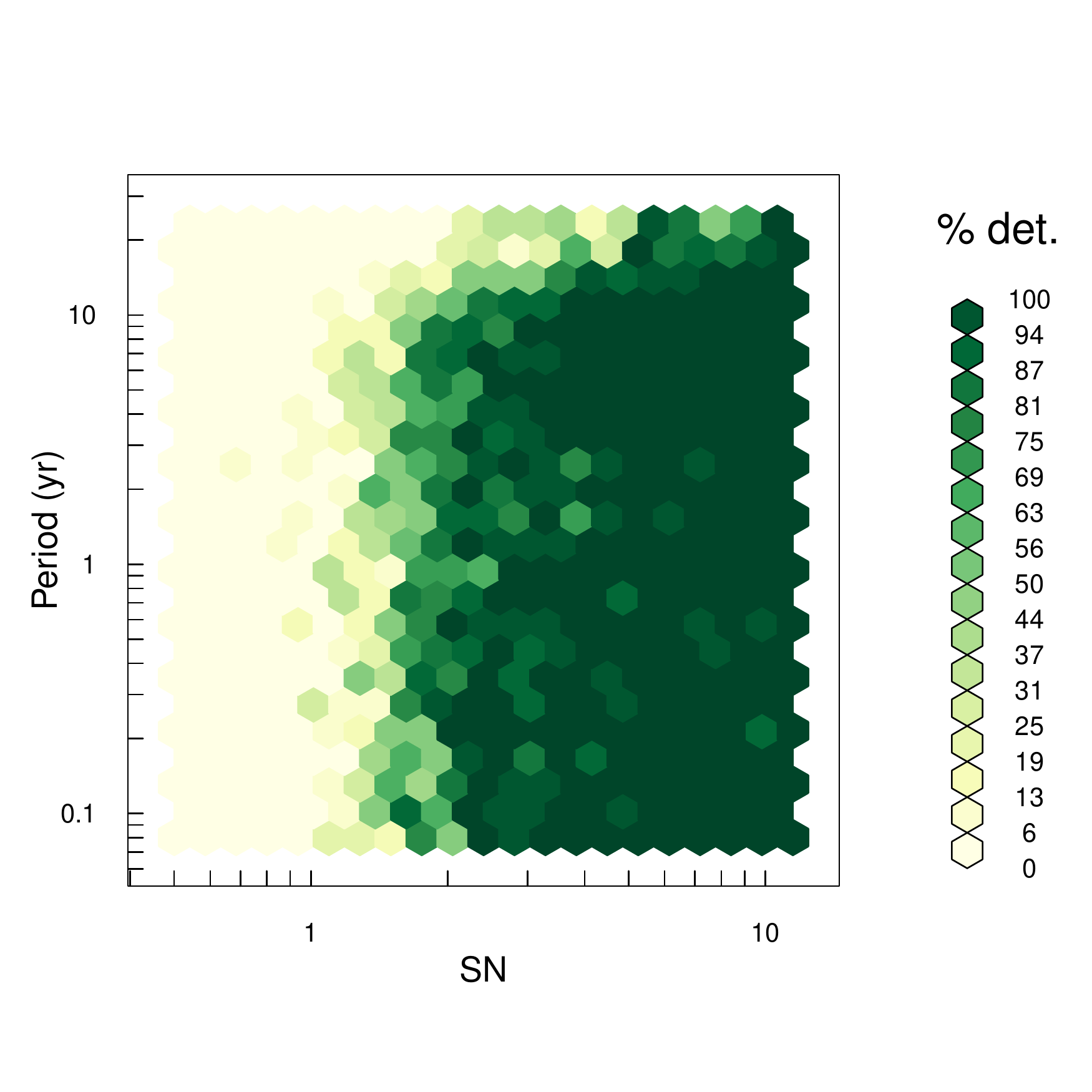}
  \caption{Detection fractions as a function of S/N and period. Left
    panel: 5~yr mission; right panel: 10~yr mission. The colours show
    the detection fraction and range from light yellow (low detection
    fraction) to dark green (high detection fraction). The transition
    between non-detections and detections is sharper and occurs at
    lower S/N in the 10~yr mission than in the 5~yr mission. Planets with
    periods longer than the mission length require a stronger S/N to be
    detected than planets with lower periods.
 }
  \label{fig:detFract2d_v8}
\end{figure*}

In Fig.~\ref{fig:detFract2d_v8} we show the detection fraction as a
function of both S/N and period.
The fraction of detected systems depends mainly on the S/N. Detecting
systems with periods longer than the mission length requires a S/N that
is higher by an amount that increases with the period.

The 50\% detection rate for systems with period shorter than the mission
length is found at S/N$\sim 2.3$ and 1.7 for the 5~yr and 10~yr
mission, respectively. Starting the MC with the circular orbit model
or at the ideal point makes little difference because the  detection rates
vary by less than 10\%.

The above numbers only consider systems for which the MC achieved
convergence within the allowed number of samples, according to the
criterion described in Sect.~\ref{sec:chain-convergence}. We found
that the fraction of non-converging chains depends on the starting
point: in the ideal starting point case, 99\% of chains converged,
while in the case of the circular orbits starting point, the fraction
lowers to $\sim 80\%$. This fraction might increase by using a more
refined method to select a starting point (as was done in e.g.
\citealt{sahlmann2013}).

\section{Parameter recovery}
\label{sec:param-recovery}

In this section\ we  consider candidate detections, and look at how
well the orbital parameters are recovered. The results depend on how
close the initial point for the MC is to the true value of the
parameters. To show the impact of the MC starting point, we 
present the results for the ideal case (starting from true values) for
period and eccentricity. We  then show, for all orbital
parameters, the results obtained by starting the MC at the best-fit
points obtained using the circular orbits model.

For every parameter, we calculated the median (the circular median is
used for $\Omega$, $\omega$, and $T_p$; see \citealt{pewsey2013}) and
an estimate of the associated error: for $\Omega$, $\omega$, and $T_p$
we used the circular standard deviation\footnote{Defined as
  $\hat\sigma=\sqrt{-2 \ln \bar R}$, where
  $\bar R = \frac{1}{n} \sum_{i=1}^n \cos(\gamma_i - \bar\gamma)$ is
  the sample mean resultant length, for a sample of $n$ of angles
  $\gamma_i$ whose mean is $\bar \gamma$, and calculated using the
  \texttt{sd.circular} function in the \texttt{circular} package for
  the R software; see \citet{pewsey2013}.}; for every other parameter,
we used the 68.3\% highest posterior density interval (HPDI) from the
posterior samples. These quantities are plotted in
Figs.~\ref{fig:PP_v8}--\ref{fig:WW_v8}.

\subsubsection*{Period}

\begin{figure*}
  \centering
  \includegraphics[width=.9\columnwidth]{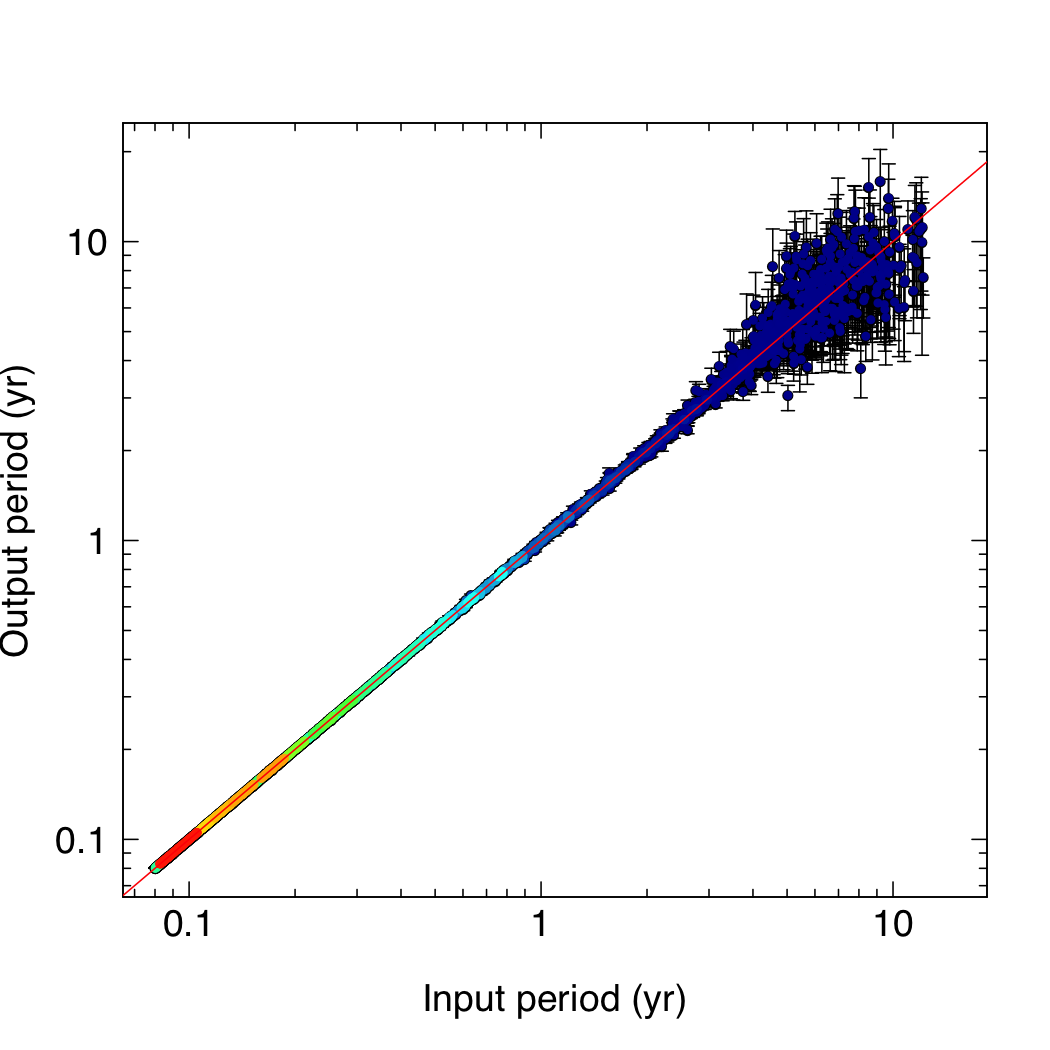}
  \includegraphics[width=.9\columnwidth]{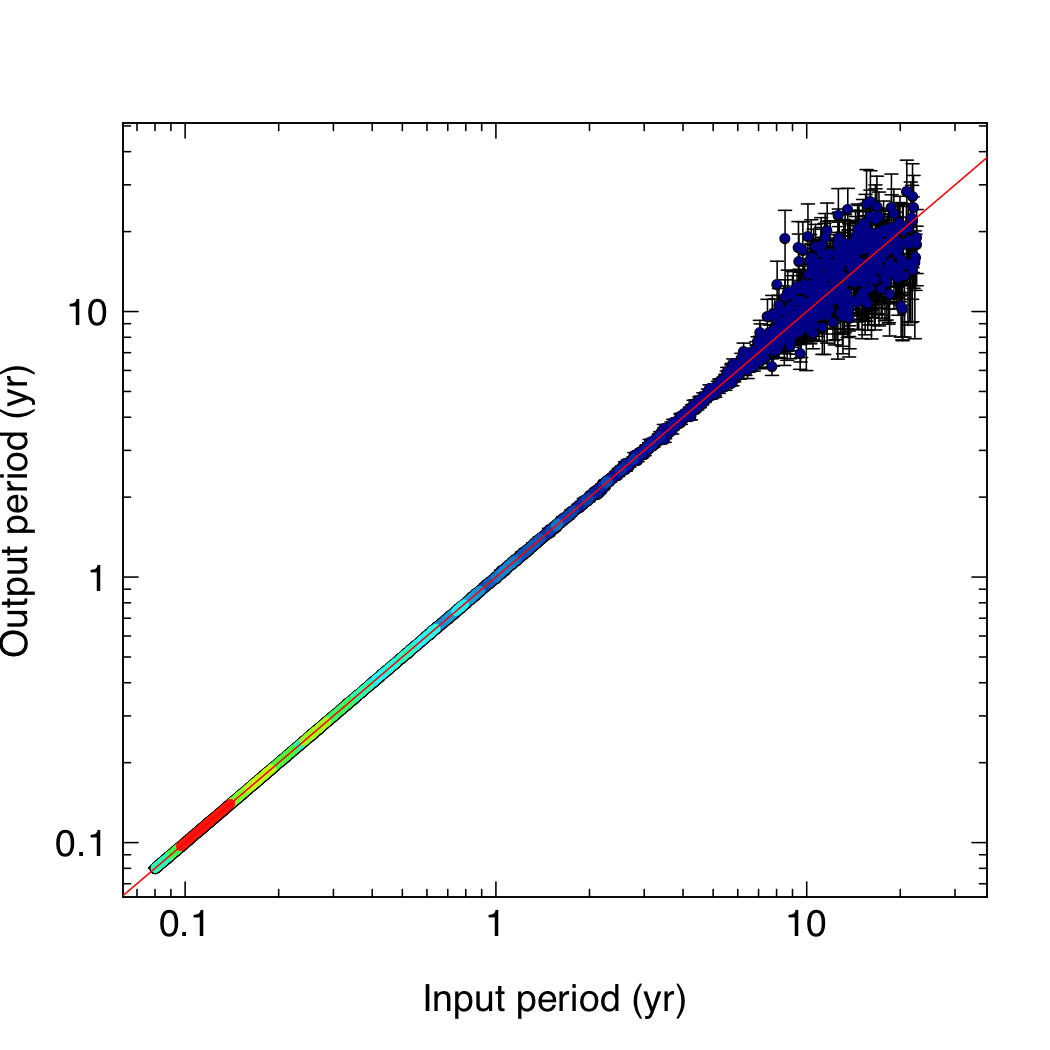}\\
  \includegraphics[width=.9\columnwidth]{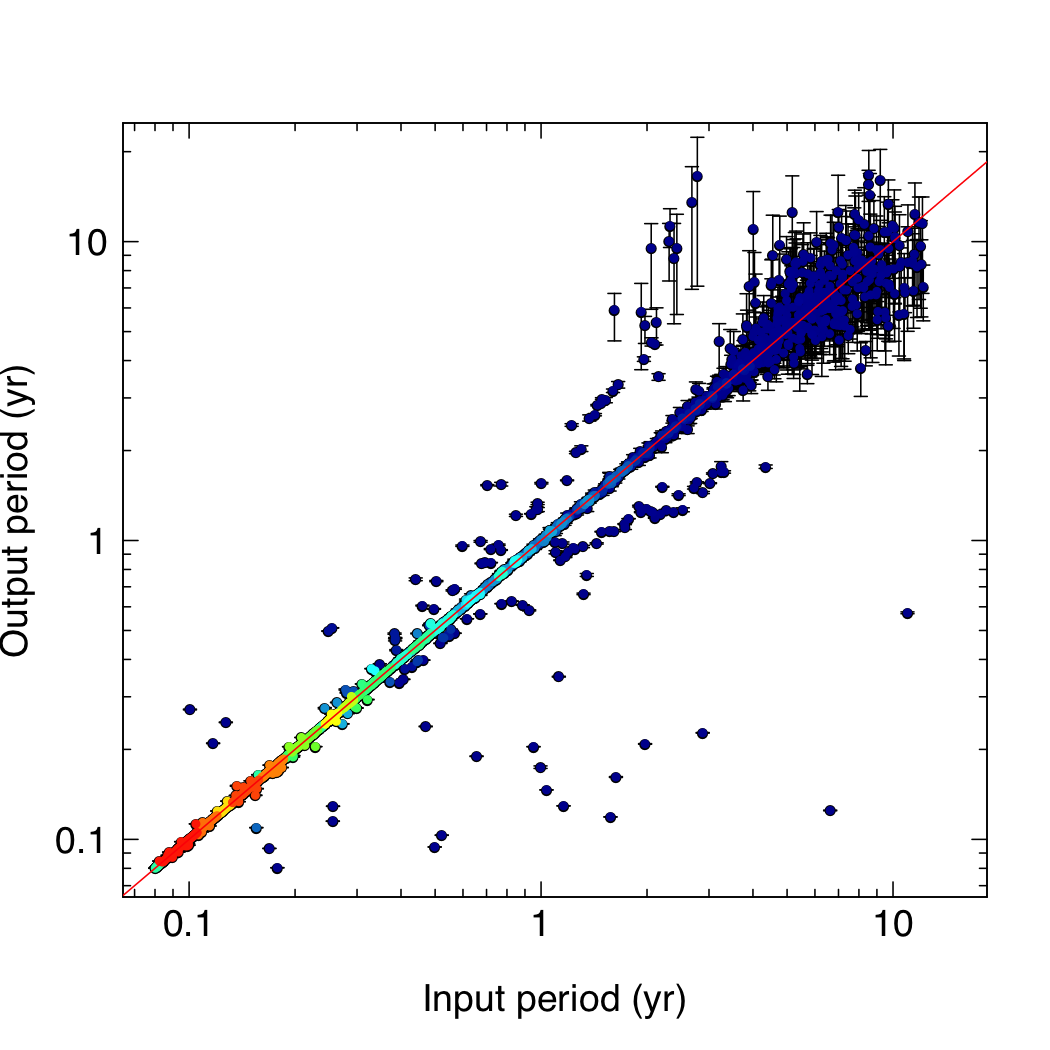}
  \includegraphics[width=.9\columnwidth]{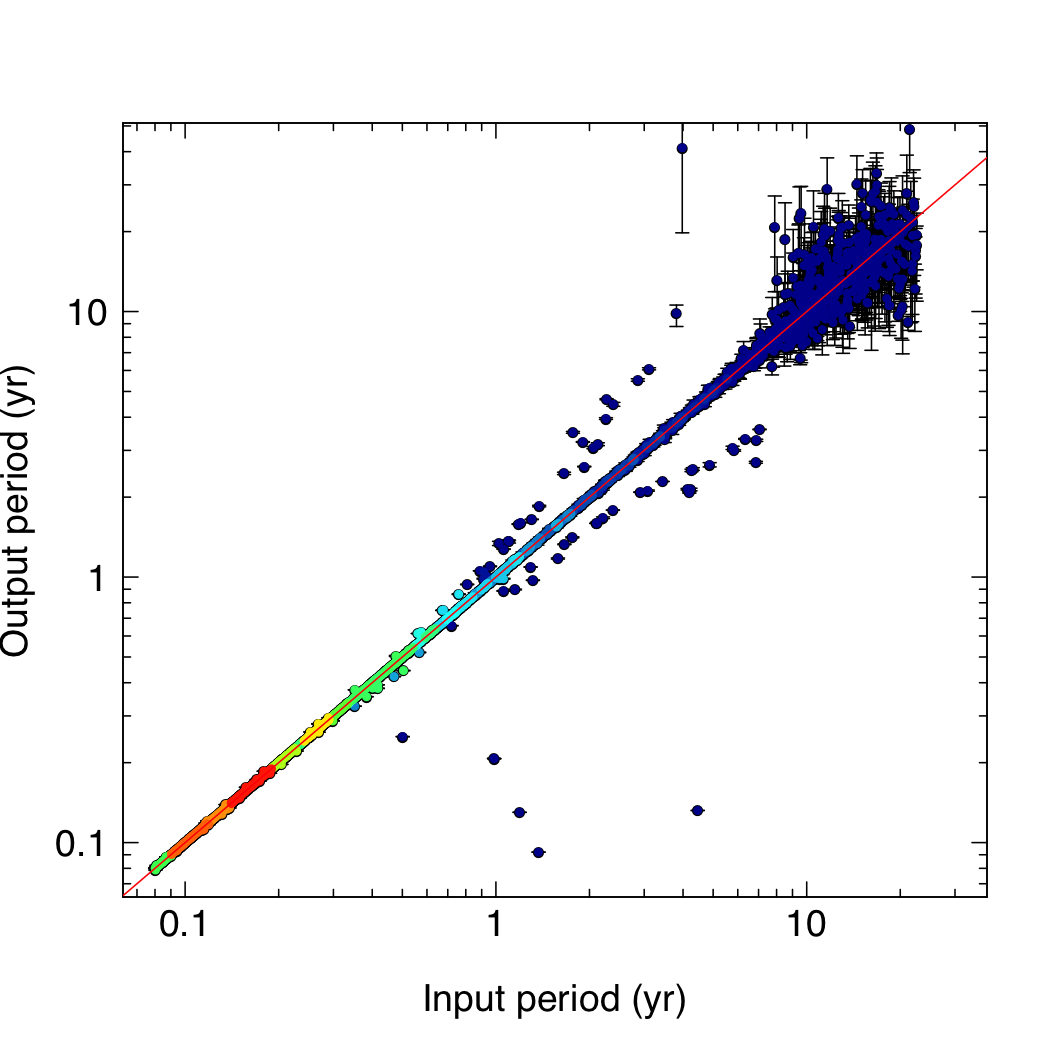}
  \caption{Simulated vs.\ recovered period. The left and right panels
    show the 5~yr and 10~yr missions, respectively. The upper
      and lower panels differ only for the starting point of the MC:
      the ideal case (MC started at the true value) in the upper
      panel, and the circular-orbit model best-fit point in the lower
      panel.  The colours show the density of points where they
    overlap, going from blue (no overlap) to red (maximum
    overlap). The error bars show the 68.3\% highest posterior density
    interval. The red solid lines show the one-to-one
    relationship. The period is correctly recovered in the
      majority of cases. The two sequences parallel to the main
    relationship are populated with systems that are incorrectly fit
    with edge-on, high-eccentricity orbits.  }
  \label{fig:PP_v8}
\end{figure*}

Figure~\ref{fig:PP_v8} (upper panels) shows that the periods are almost
always correctly recovered in the ideal case, when the MC are started
using the true values of the parameters.

Instead, the use of imperfect starting points produces a small number
of non-recovered periods (the systems with
$0.1\lesssim P\out\lesssim 0.3$~yr). Also, a small number of systems
are fit with periods that are somewhat shorter or longer than the
input value by a factor of $\sim 40$--$60\%$ (they are visible in
Fig.~\ref{fig:PP_v8} as the two sequences right above and below the
one-to-one line); these systems are also mis-fit as edge-on,
high-eccentricity systems (they have incorrect $\cos i\out \sim 0$ and
$e\out\sim 1$). The fraction of such systems depends on the mission
length but not on the S/N and is, considering only inputs with
$0.2\le P\inp\le 3.5$~yr, 12\% for the 5~yr mission and 3.2\% for the
10~yr mission. The two sequences can be interpreted as parallel
sequences in terms of frequency $f=1/P$, where the frequency of each
sequence is $f\out=f\inp \pm 0.3$.

\subsubsection*{Eccentricity}

\begin{figure*}
  \centering
  \includegraphics[width=.9\columnwidth]{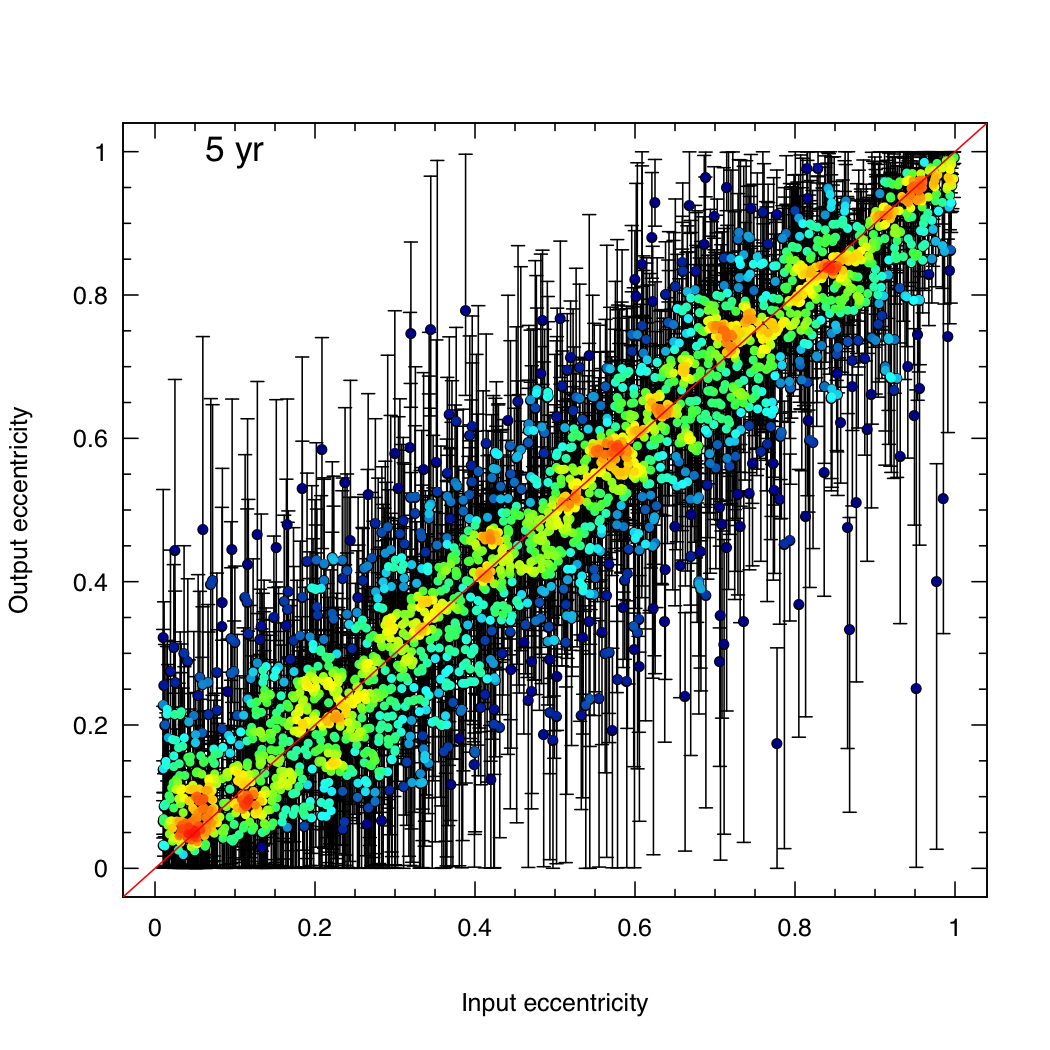}
  \includegraphics[width=.9\columnwidth]{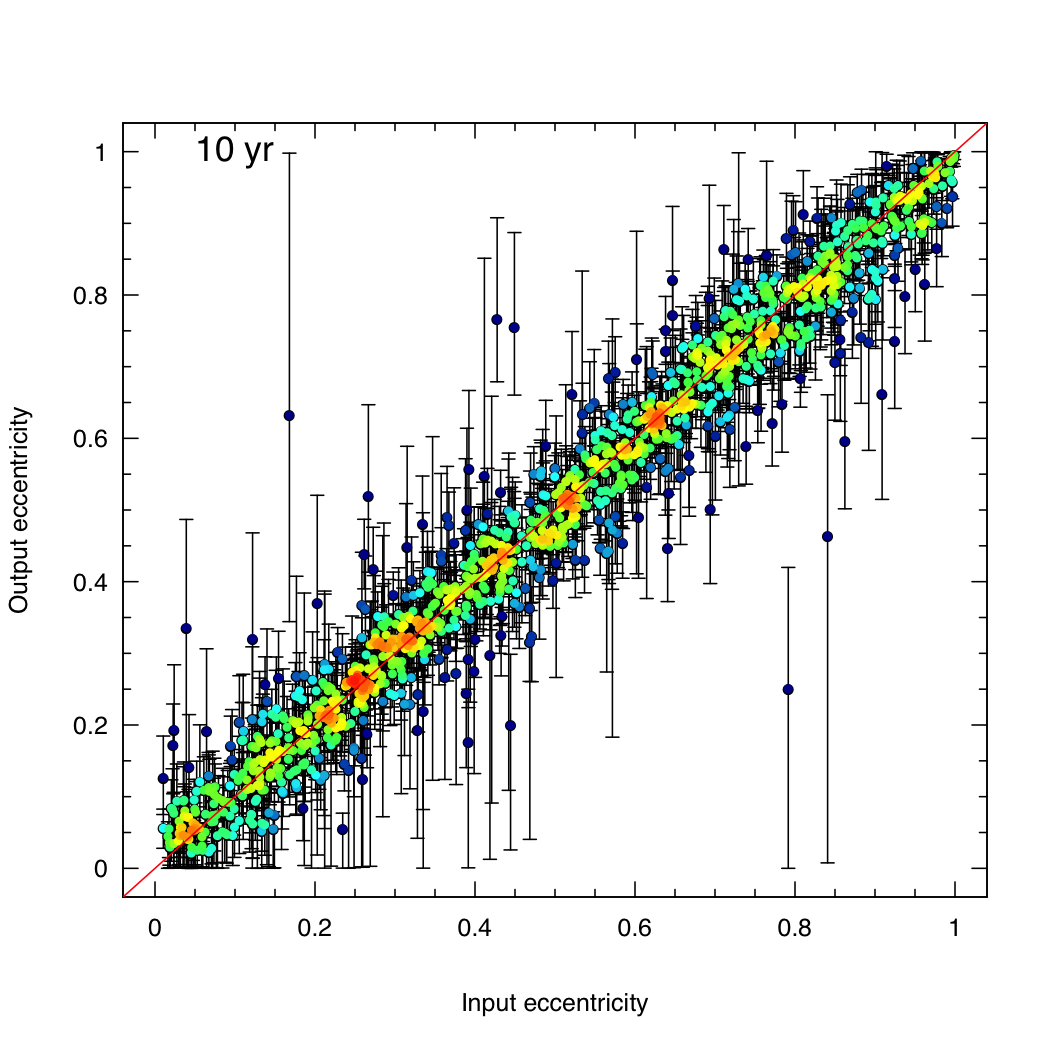}\\
  \includegraphics[width=.9\columnwidth]{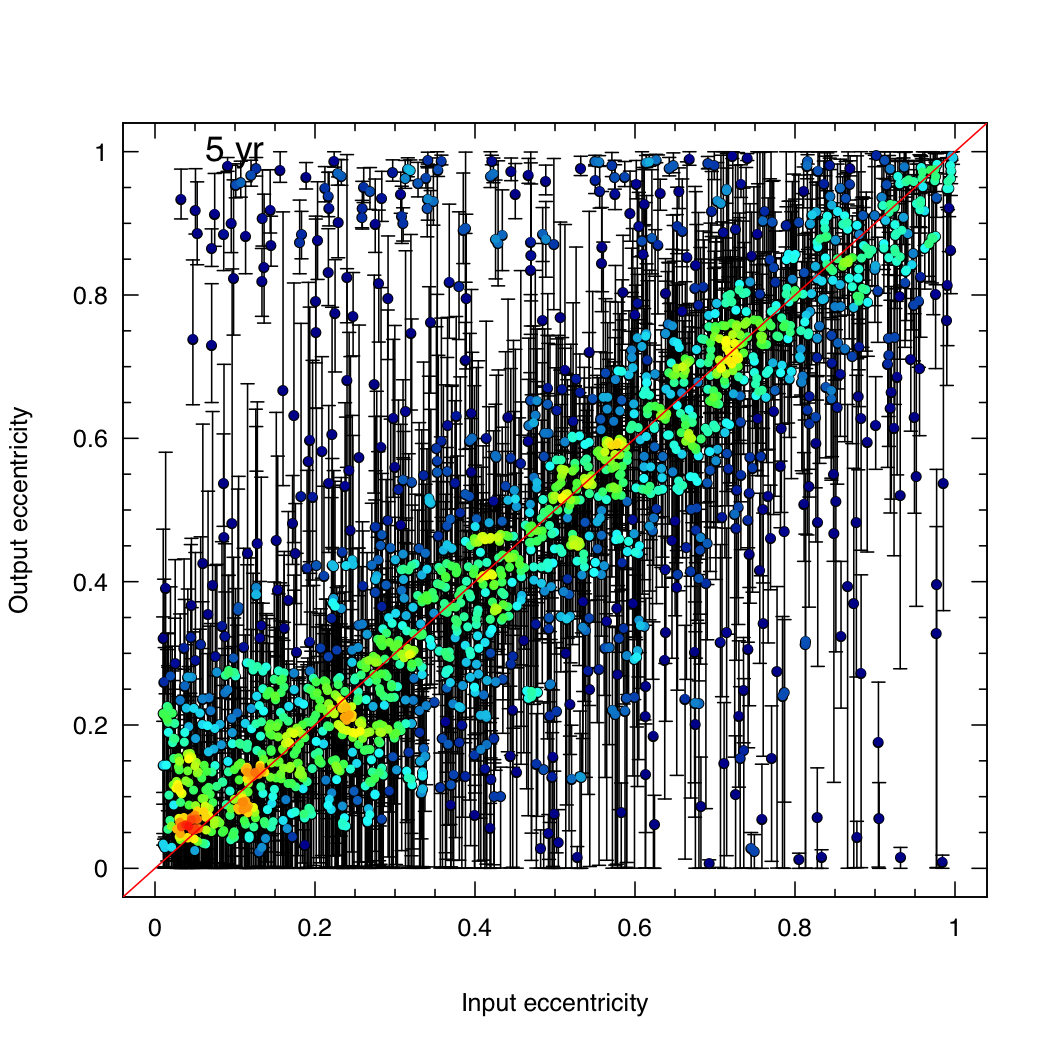}
  \includegraphics[width=.9\columnwidth]{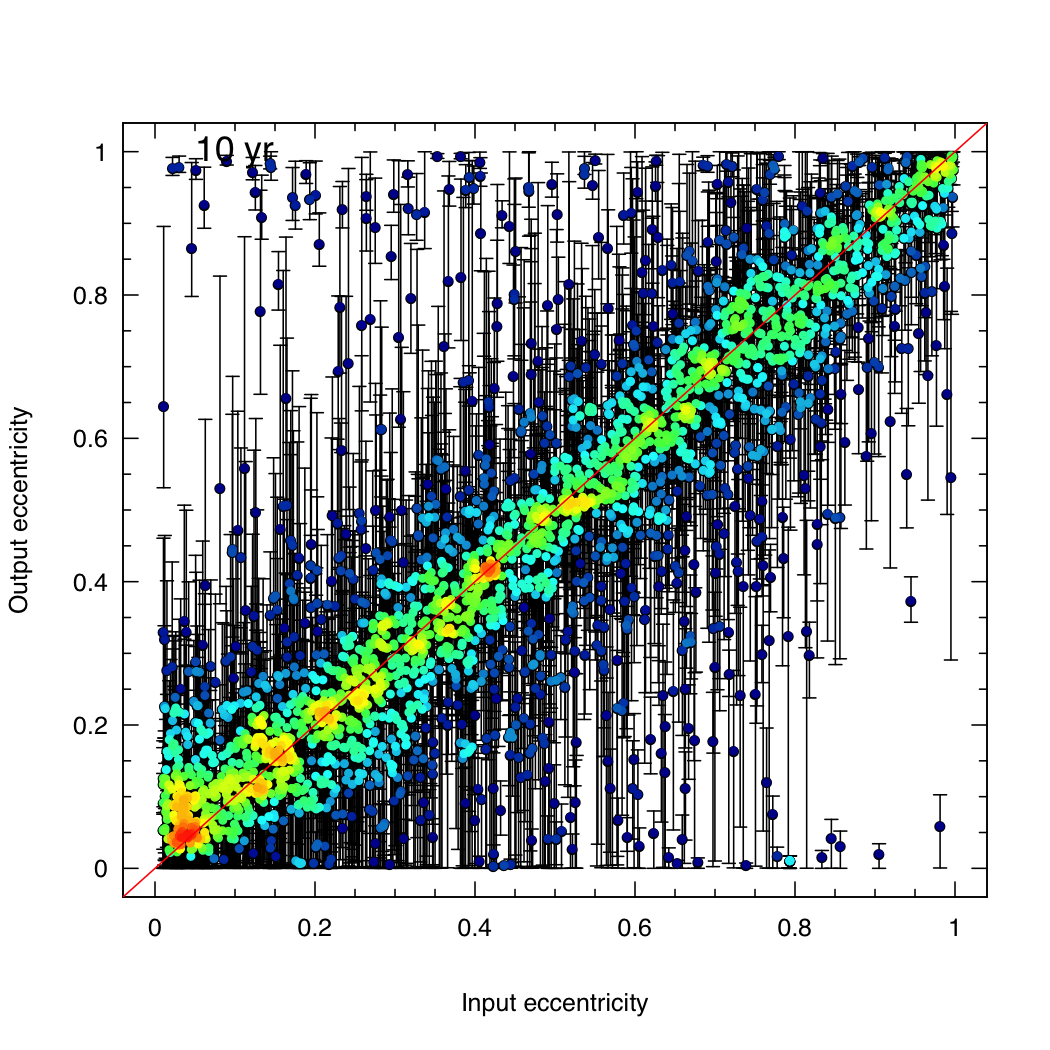}
  \caption{Simulated vs.\ recovered eccentricity. Panels, symbols, and
    colours as in Fig.~\ref{fig:PP_v8}. The eccentricity is
      recovered in most cases, albeit with large uncertainties.  When
      the MC are started using the circular orbits model (lower
      panels), some
      systems with $e\inp\lesssim0.4$ are incorrectly fit with very elliptical orbits
      ($e\out \sim 1$), especially in the 5~yr mission; these systems
      are the same as those that lie in the parallel sequences of
      Fig.~\ref{fig:PP_v8}.}
  \label{fig:ee_v8}
\end{figure*}

The eccentricity exhibits quite a large scatter, both in terms of the
location of the median $e\out$ and of the size of the HPDI. The
scatter is somewhat larger when the MC are started using the circular
orbit model.

When the MC are started using the circular orbit model, a fraction of
systems are fit with $e\out\sim 1$ with no dependence on $e\inp$; this
happens in 9.6\% and 6.7\% of the systems with $e\inp<0.8$ for the
5~yr and 10~yr mission, respectively. Such systems also exhibit
incorrect fits in other variables: they have
$\omega\out\sim\pm90^\circ$, an incorrect $T_p$, and
$\cos i\out \sim 0$ (edge-on); their $P\out$ also run on the parallel
sequences to the one-to-one relationships.

\subsubsection*{Size of the orbit}

\begin{figure*}
  \centering

  \includegraphics[width=.9\columnwidth]{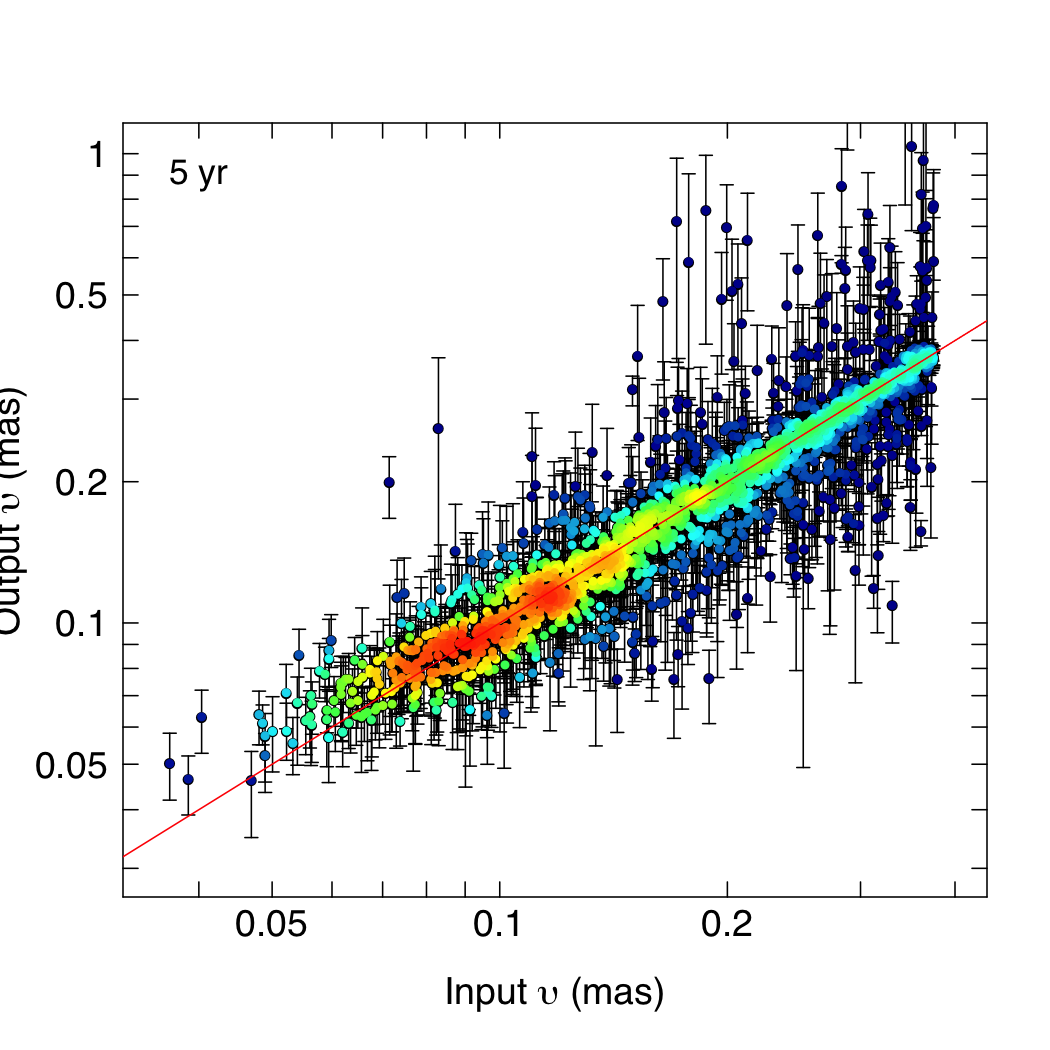}
  \includegraphics[width=.9\columnwidth]{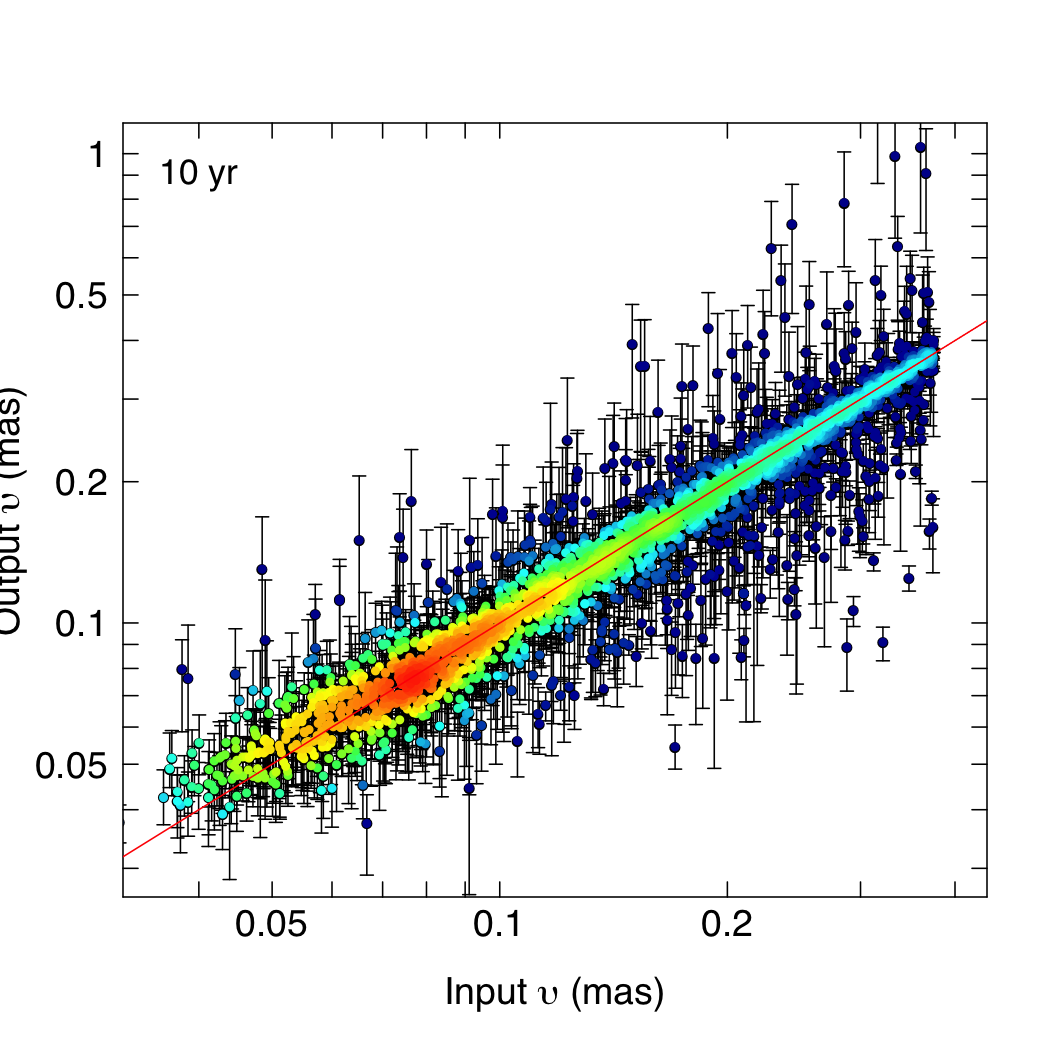}
  \caption{Simulated vs.\ recovered angular size of the
    orbit. Symbols and colours as in Fig.~\ref{fig:PP_v8}. The
      two panels show only the results when the MC have been started
      using the circular orbit model. Most of the points with
    $\upsilon\out \ll \upsilon\inp$ have input periods longer than the
    mission length, and form the horizontal plume at large $P\inp$ in
    Fig.~\ref{fig:PP_v8}.  }
  \label{fig:as_v8}
\end{figure*}

The angular size of the semi-major axis of the orbit is given by the
astrometric signature $\upsilon$; its input value $\upsilon\inp$ can
be obtained from the S/N through Eq.~(\ref{eq:SN}). In
Fig.~\ref{fig:as_v8}, we show that the one-to-one relationship is well
recovered.

Conversely, most of the systems with deviations on the left side
(i.e.\ with $\upsilon\out > \upsilon\inp$) are also mis-fit as
edge-on, high eccentricity systems ($\cos i\out \sim 0$ and
$e\out\sim 1$).

\subsubsection*{Inclination}

\begin{figure*}
  \centering

  \includegraphics[width=.9\columnwidth]{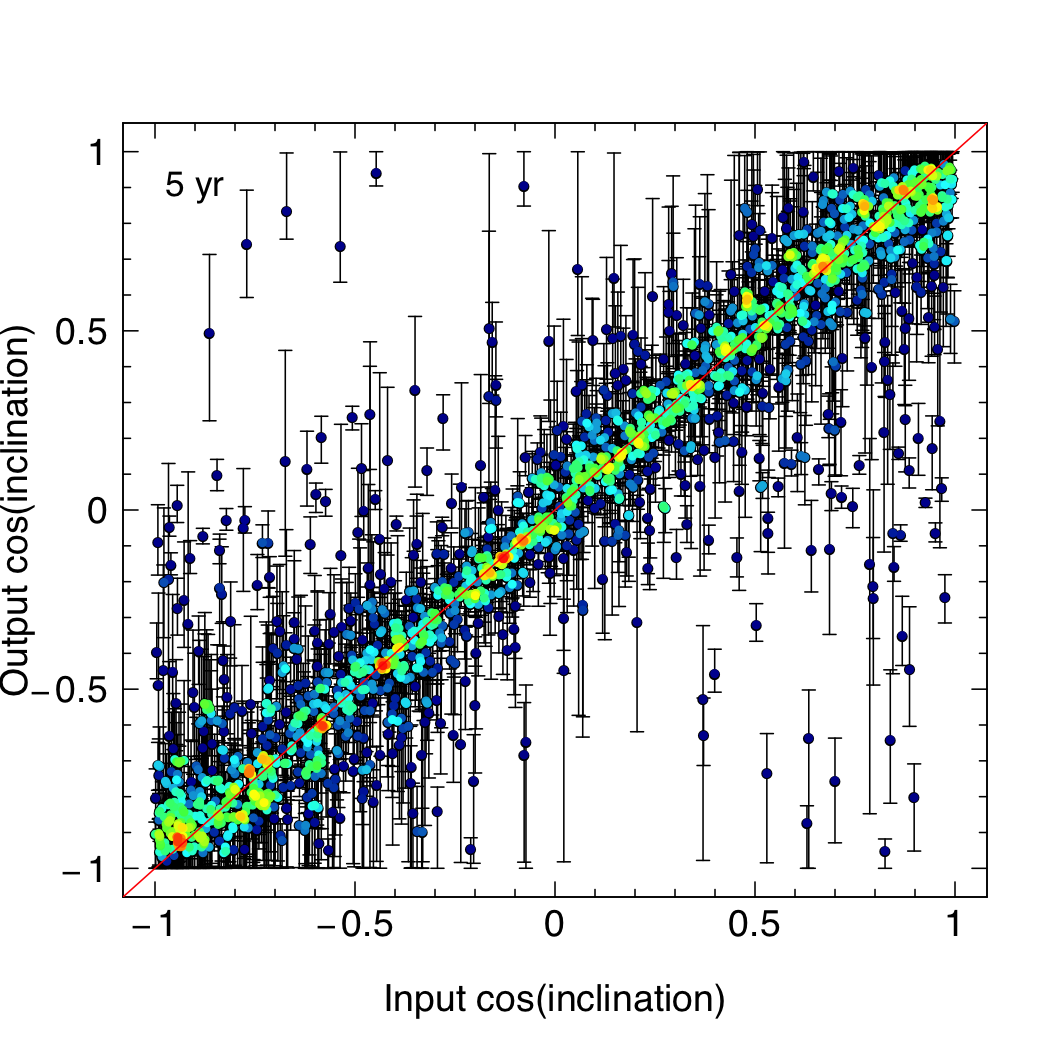}
  \includegraphics[width=.9\columnwidth]{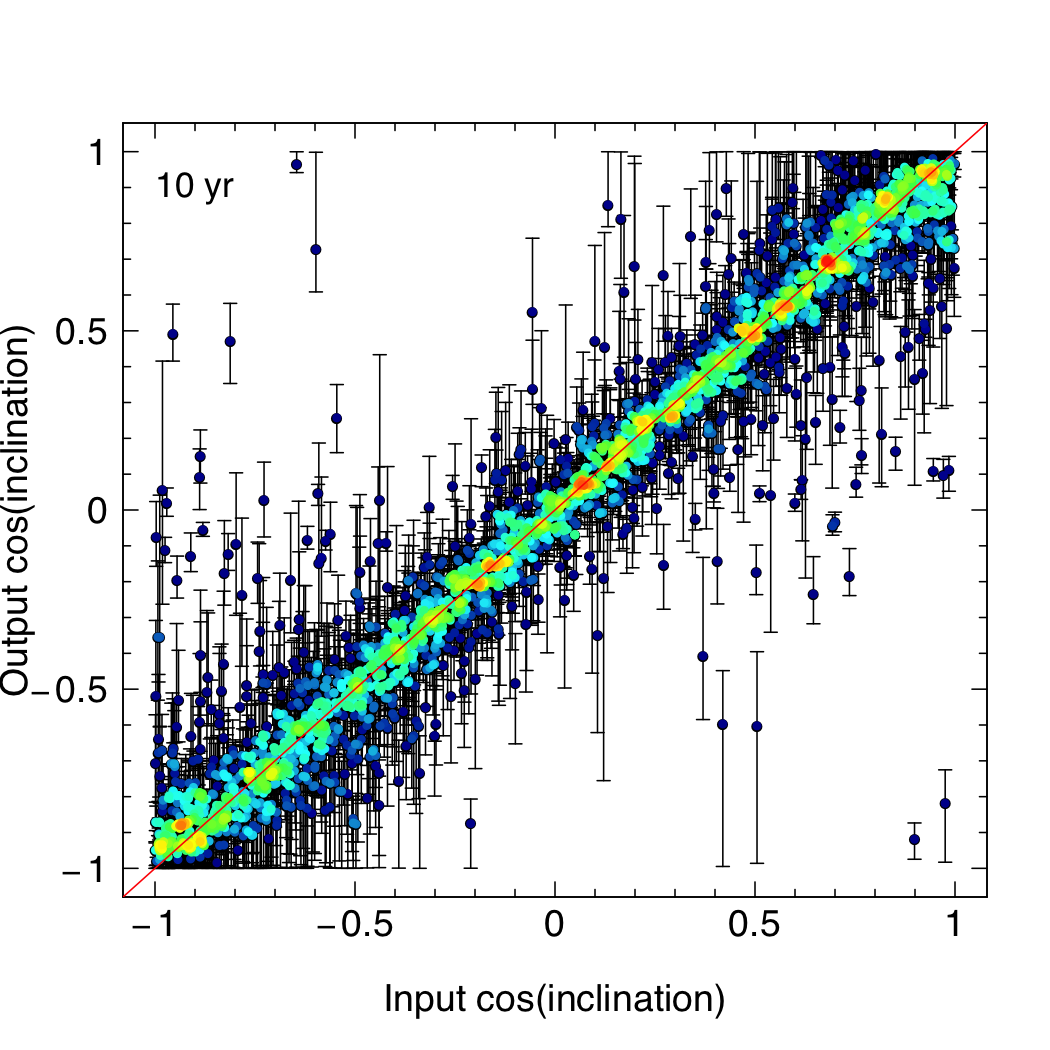}
  \caption{Simulated vs.\ recovered inclination. Panels, symbols, and
    colours as in Fig.~\ref{fig:as_v8}. The inclination is correctly
    recovered for most systems. Especially in the 5~yr mission, there
    is a fraction of systems which are mistaken as edge-on
    ($\cos i\out \sim 0$) which seems to be independent of the input
    inclination, but which is often associated with high eccentricity
    ($e\out \sim 1$) and with the parallel sequences in the period.}
  \label{fig:ii_v8}
\end{figure*}

The inclination is recovered correctly in the majority of cases
(Fig.~\ref{fig:ii_v8}); the scatter of the one-to-one relationship is
reduced when going from the nominal to the extended mission.

\subsubsection*{Longitude of ascending node}
\label{sec:om_b}

\begin{figure*}
  \centering

  \includegraphics[width=.9\columnwidth]{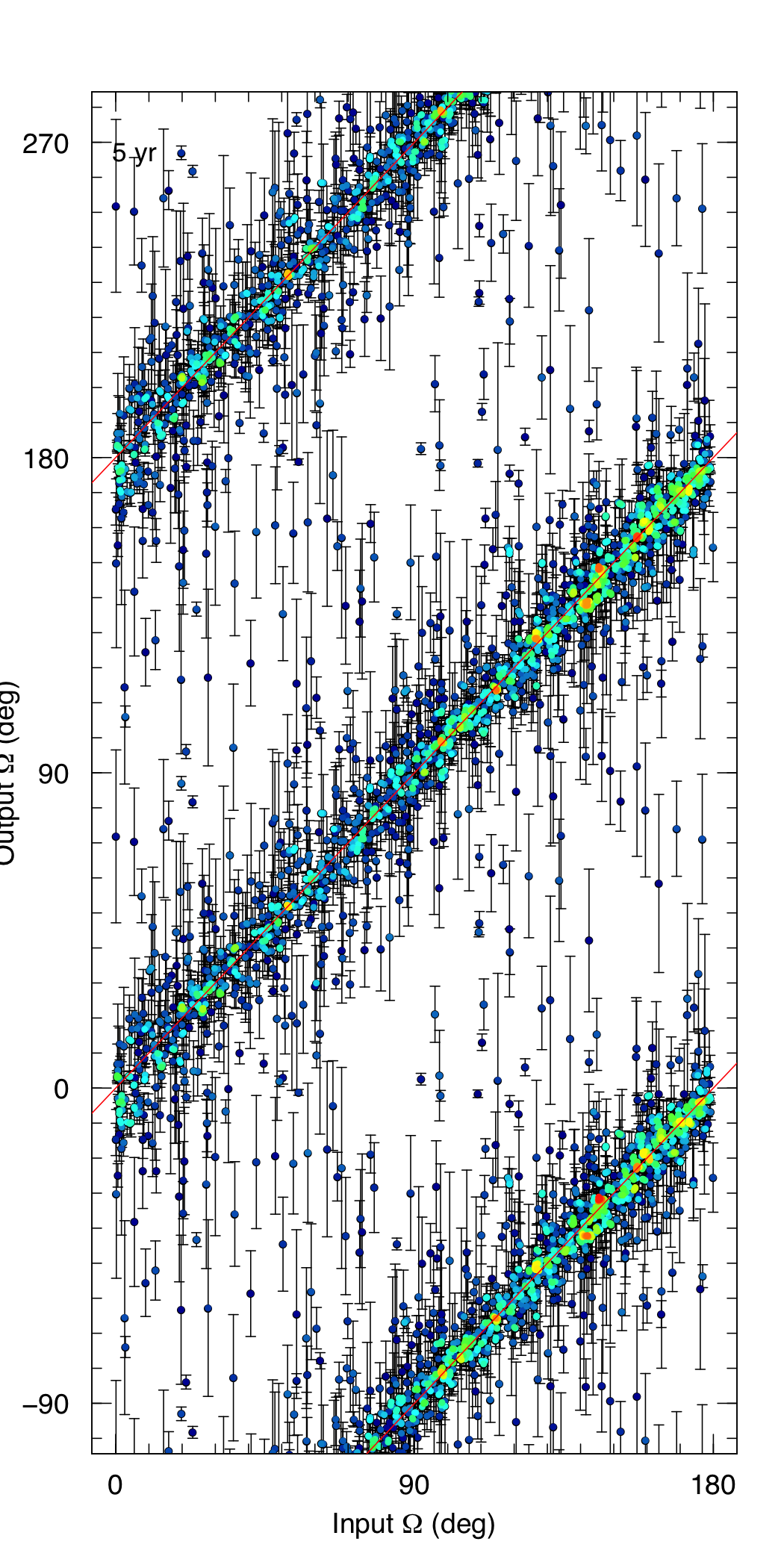}
  \includegraphics[width=.9\columnwidth]{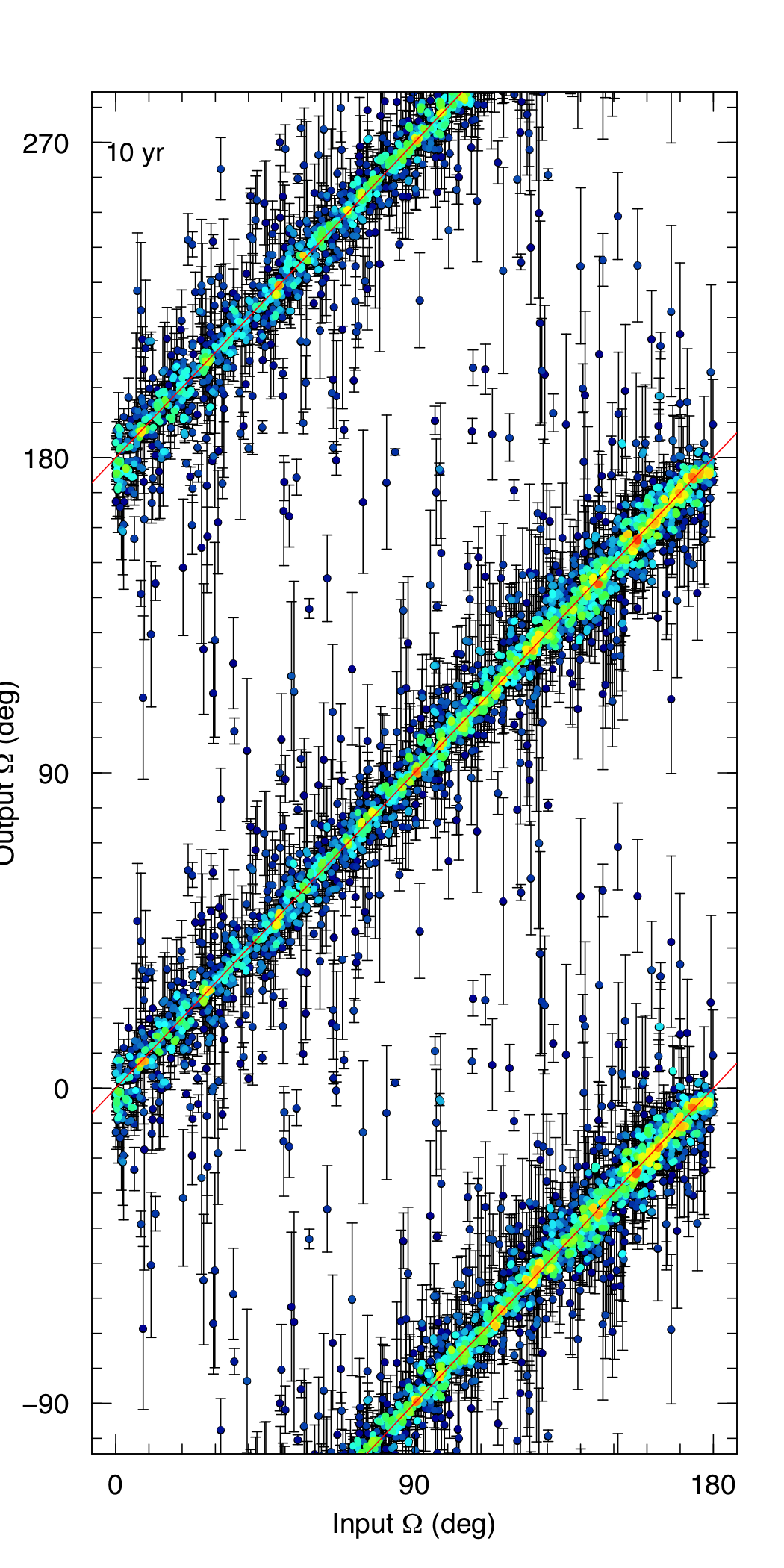}
  \caption{Simulated vs.\ recovered longitude of ascending node
    ($\Omega$). Panels, symbols, and colours as in
    Fig.~\ref{fig:as_v8}. Since $\Omega$ is an angle, we plot the
    data points twice (at nominal position and at $\pm 180^\circ$) to
    better show the structure of the noise.}
  \label{fig:WW_v8}
\end{figure*}

The longitude of the ascending node ($\Omega$) is correctly recovered
in most cases; see Fig.~\ref{fig:WW_v8}. There does not seem to be any
particular structure in the points that deviate most from the
one-to-one relationship.

\subsubsection*{Argument of periapsis}
\label{sec:om_s}

Whether the argument of periapsis ($\omega$) is correctly recovered
depends on the eccentricity. As discussed in Sect.~\ref{sec:chi2},
when orbits are nearly circular, $\omega$ loses its geometrical
meaning and starts to track the zero-point of the orbit as
$T_p$. Therefore, for low eccentricity the relationship between
$\omega\inp$ and $\omega\out$ is lost.

The relationship is recovered for larger eccentricities $e\inp\gtrsim 0.1$;
the relationship for $\omega$ appears somewhat noisier than the
analogous relationships that hold for $\Omega$ and $T_p$.

\subsubsection*{Periapsis transit time}
\label{sec:tpn}

As already seen in the case of $\omega$, the periapsis transit time $T_p$ is only
recovered when orbits are not circular.
For systems with $e\inp>0.1$, we found that  the one-to-one
relationship seems well recovered, with no particular structure in the
points deviating from it.

\section{Discussion}
\label{sec:discussion}

\subsection{Maximum distance of detectable planets}
\label{sec:maxdistance}

Previous studies (C08, P14) had suggested a minimum $\SN=3$ for the
detection of planets. This corresponds
to a 74\% chance of detecting a planet in the 5~yr mission (92\% for
the 10~yr mission). A 50\% detection chance occurs at $\SN=2.3$ and
1.7, respectively (Sect.~\ref{sec:detectionrate}).

We can therefore take a second look at Fig.~\ref{fig:sn-vs-d}, where
we  present the S/N of a few representative classes of planets, as
a function of their distance from the solar system. Assuming the S/N
that allow a 50\% detection chance, the maximum distance up to which a
Jupiter-mass planet around a 1~$M_\sun$ star with a semi-major axis of
1 au could be detected is 13 pc for the 5~yr mission, and 17 pc for
the 10~yr mission. Considering a semi-major axis of 3 au, the
distances would increase to 39 pc and 52 pc, respectively. A 10~yr
mission also allows the detections of planets with semi-major axes up
to 4~au, yielding a maximum distance of 70 pc.

For a Neptune-mass planet, the distances are shorter. Even in the
favourable case of a 3 au semi-major axis, the distances would be 1.9
and 2.6 pc, respectively. For the 10~yr mission, a 4 au semi-major
axis would yield a distance of 3.5 pc. In all cases, there is a better chance to
detect planets   if one also considers stars with a lower mass
than the 1~$M_\sun$ considered here.

In a real survey, it is possible that some false positives will be
included among the detections of planets with low S/N. The
$\Delta\mathrm{BIC}=20$ threshold that we have considered has allowed
zero false positives among our no-planet simulations; therefore,
considering the number of simulated systems, upper limits to the false
positive fraction can be put as $<1.5\e{-4}$ and $<2.3\e{-4}$ for the
5 yr and and 10 yr mission, respectively.  To put such numbers in the
context of a real survey, we consider the number of stars within 100
pc of the Sun. Considering  only dwarf stars of the F, G, and K
spectral types (as done previously by C08 and P14), the Gaia universe
model snapshot \citep[GUMS,][]{robin2012-gums} includes $6.3\e{4}$
stars. Therefore, from the above upper limits to the false positive
fraction, one could estimate that $\lesssim 9$ and $\lesssim 14$ false
positives could be present in an all-sky survey.

\subsection{Accuracy and precision of fit parameters}
\label{sec:errors}

\begin{table}
  \caption[]{Distribution of the accuracy with which fit
    parameters are recovered.  The first column
    presents the orbital parameter; the other columns present the
    distribution of the length of its 68.3\%
    error bars in the 5~yr mission (Cols. 2--4) and
    in the 10~yr mission (Cols. 5--7) as the 25\%
    quartile, median, and 75\% quartile.  Relative errors are shown for quantities that vary
    over several orders of magnitude, namely $P$ and $\upsilon$; absolute errors for
    the remaining parameters. We consider $T_p/P$ (instead of just
    $T_p$) because it is always contained in the [0, 1] interval.}
  \label{table:accuracy}
  \centering                       
  \begin{tabular}{lrrrrrr}
    \hline\hline      
            & \multicolumn{3}{c}{5 yr} & \multicolumn{3}{c}{10 yr}                                       \\
            & 25\%                     & 50\%       & 75\%       & 25\%        & 50\%       & 75\%       \\
    \hline                       
 $P$        & 0.14\%                   & 0.70\%      & 6.5\%      & 0.058\%      & 0.29\%     & 1.8\%      \\ 
 $\upsilon$ & 3.1\%                     & 8.0\%       & 18\%       & 2.3\%       & 6.0\%       & 14\%       \\    
    \hline                                                         
 $e$        & 0.030                    & 0.077       & 0.17       & 0.025      & 0.061       & 0.14       \\    
 $\cos i$   & 0.031                     & 0.078       & 0.17       & 0.023       & 0.051      & 0.11       \\    
 $\Omega$   & 2.7$^\circ$              & 7.3$^\circ$ & 25$^\circ$ & 1.9$^\circ$ & 5.2$^\circ$ & 16$^\circ$ \\         
 $\omega$   & 8.4$^\circ$               & 21$^\circ$ & 45$^\circ$ & 6.0$^\circ$  & 18$^\circ$ & 41$^\circ$ \\         
 $T_p / P$  & 0.015                    & 0.045       & 0.16       & 0.011       & 0.032      & 0.11       \\
                             \hline
  \end{tabular}
\end{table}

\begin{table}
  \caption[]{Distribution of the precision with which the fit
      parameters are recovered. Columns as in Table~\ref{table:accuracy}.}
  \label{table:precision}
  \centering                       
  \begin{tabular}{lrrrrrr}
    \hline\hline      
            & \multicolumn{3}{c}{5 yr} & \multicolumn{3}{c}{10 yr}                                       \\
            & 25\%                     & 50\%       & 75\%       & 25\%        & 50\%       & 75\%       \\
    \hline                       
 $P$        & 0.40\%                   & 1.4\%      & 7.3\%      & 0.19\%      & 0.79\%     & 3.8\%      \\ 
 $\upsilon$ & 14\%                     & 23\%       & 35\%       & 9.0\%       & 17\%       & 30\%       \\    
    \hline                                                         
 $e$        & 0.12                    & 0.19       & 0.28       & 0.094       & 0.16       & 0.26       \\    
 $\cos i$   & 0.12                     & 0.19       & 0.29       & 0.087       & 0.15       & 0.25       \\    
 $\Omega$   & 9.2$^\circ$              & 17$^\circ$ & 37$^\circ$ & 6.7$^\circ$ & 13$^\circ$ & 30$^\circ$ \\          
 $\omega$   & 18$^\circ$               & 32$^\circ$ & 54$^\circ$ & 15$^\circ$  & 29$^\circ$ & 50$^\circ$ \\         
 $T_p / P$  & 0.036                    & 0.072       & 0.14       & 0.029       & 0.059      & 0.13       \\
                             \hline
  \end{tabular}
\end{table}

\begin{figure*}
  \centering
  \includegraphics[width=\columnwidth]{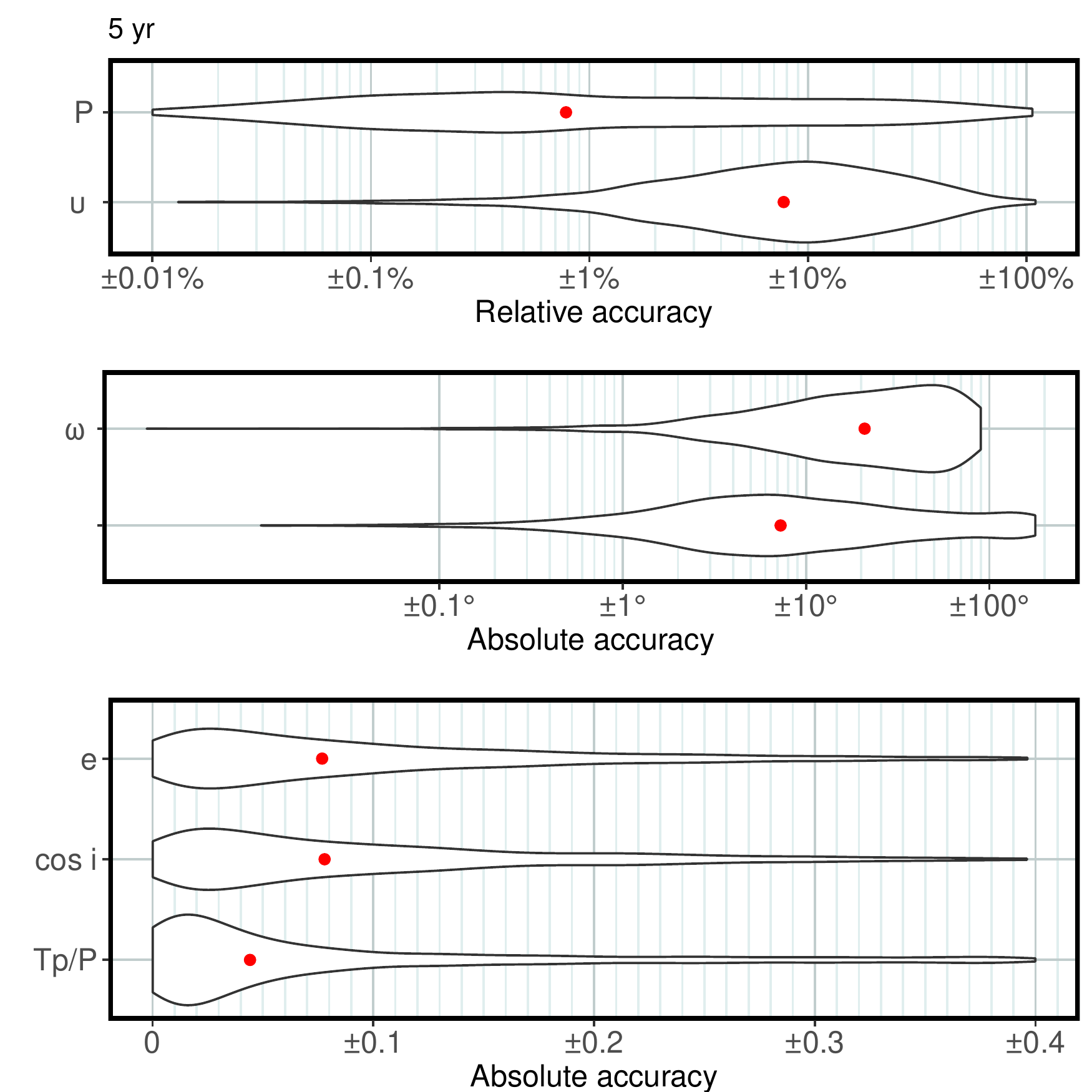}
  \includegraphics[width=\columnwidth]{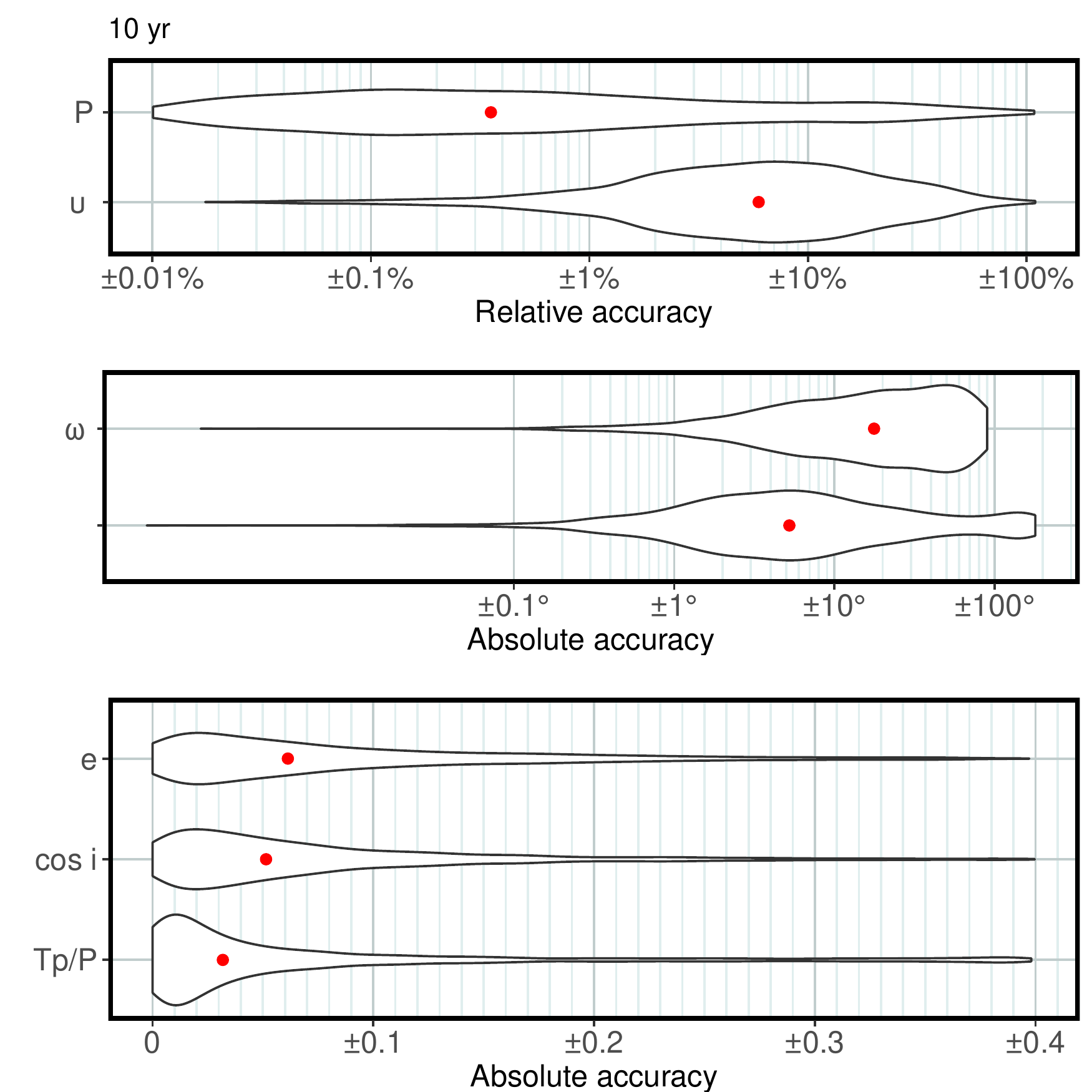}\\ \medskip
  \includegraphics[width=\columnwidth]{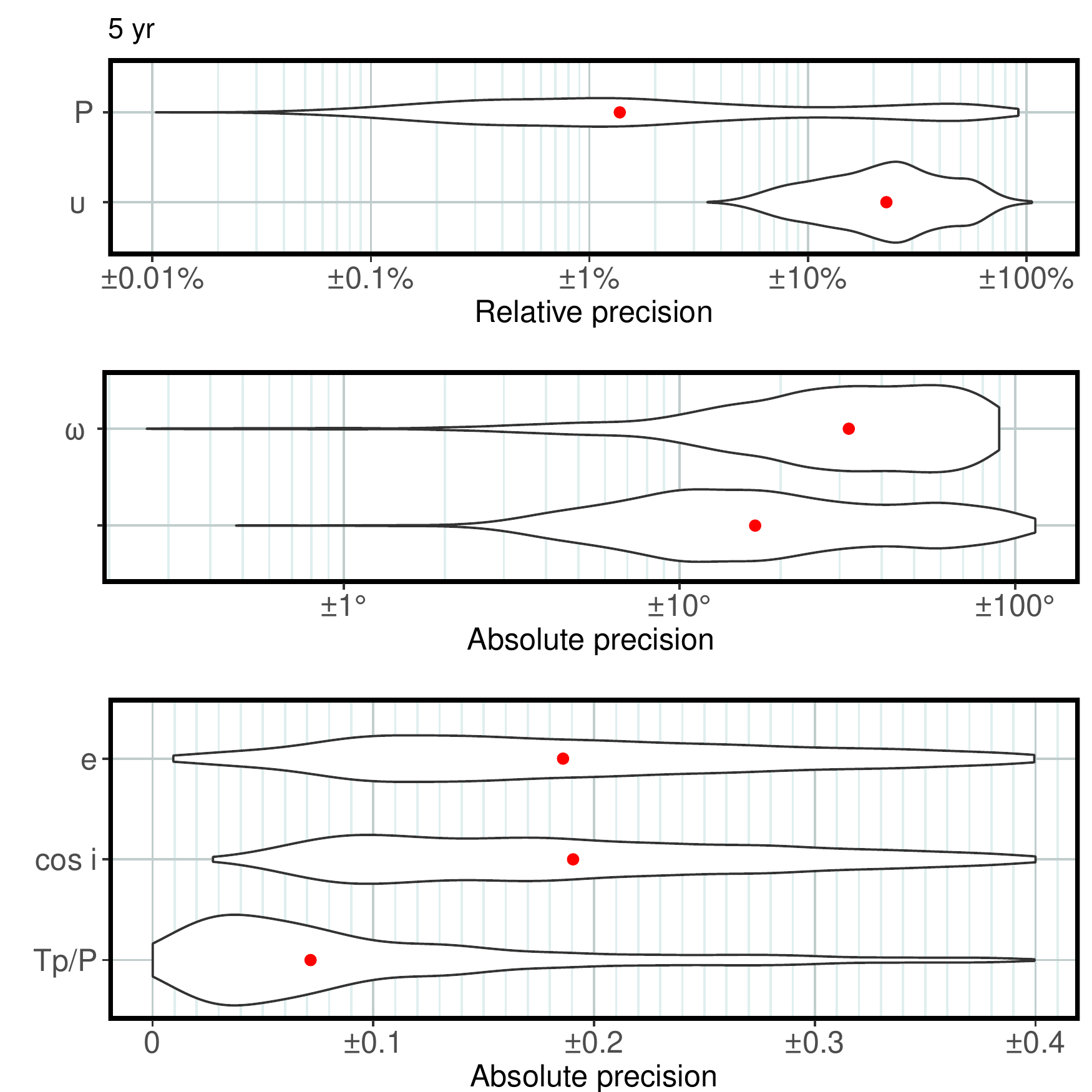}
  \includegraphics[width=\columnwidth]{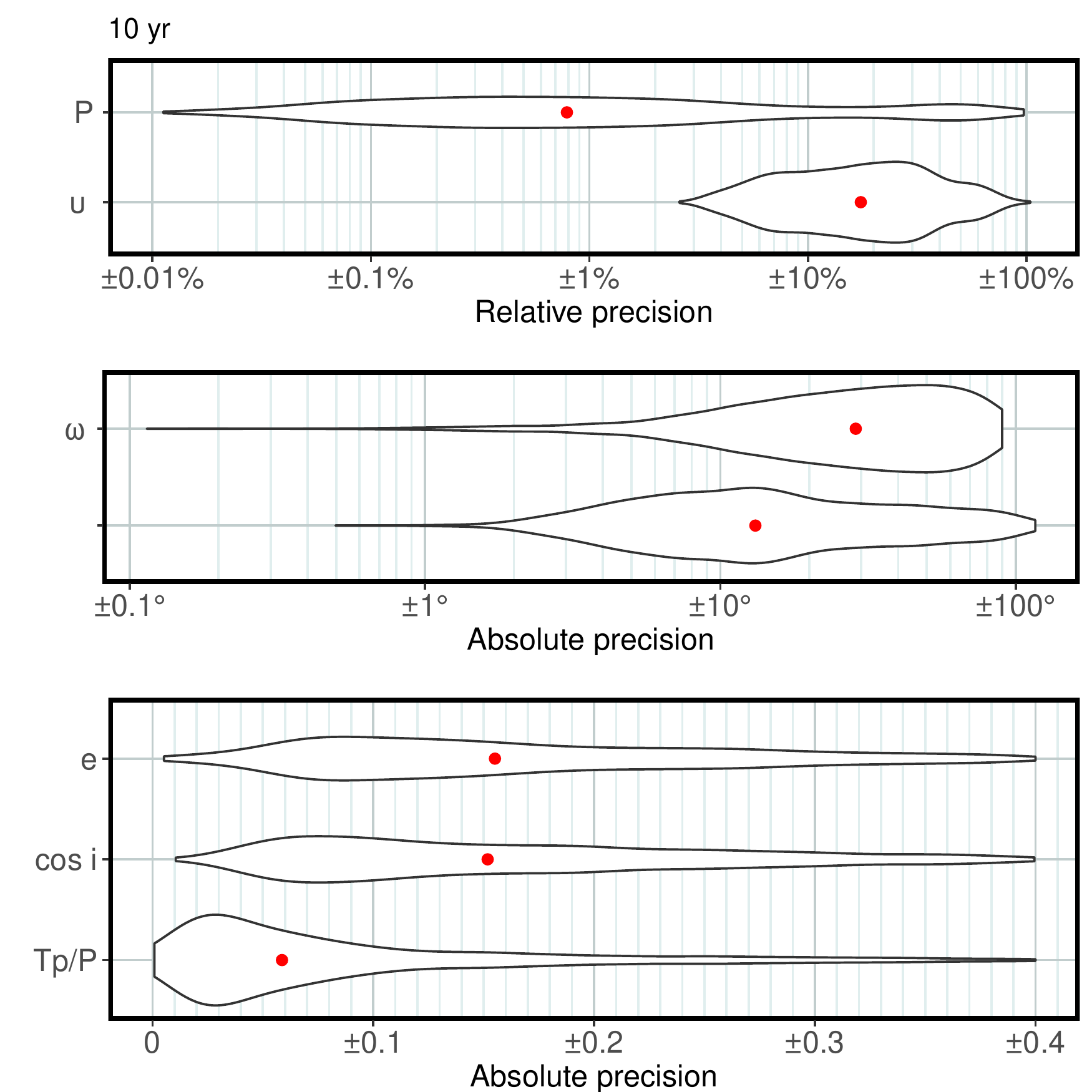}
  \caption{Accuracy and precision of the fit parameters. Left panels: 5~yr
    mission; right panels: 10~yr mission. For each orbital
    parameter, the accuracy is the absolute difference between the
    input value and the median of the output posterior
    distribution. The precision is the length of the 68.3\%
    highest posterior density interval (HPDI). The figures depicted by the black
    curves show the distribution of either accuracy or precision;
    they have normalised areas, so that at every value of the abscissa,
    their height is proportional to the density of the
    distribution. The red dots mark the
    median of each distribution.  Relative quantities are shown for quantities that vary
    over several orders of magnitude, namely $P$ and $\upsilon$; absolute quantities for
    the remaining parameters. We consider $T_p/P$ (instead of just
    $T_p$) because it is always contained in the [0, 1] interval;
    thus,  $T_p$ has a median precision of $5\% P$ in the 5~yr mission.
  }
  \label{fig:errorBarSize}
\end{figure*}

We examine which parameters are recovered better, and with what
accuracy and precision they are on average recovered. We define
accuracy as the absolute difference between the input and recovered
values, and precision as the random error that affects the recovered
values (i.e. the length of an error bar, given by the length of the
68.3\% HPDI). For the two scale parameters ($P$ and $\upsilon$, that
vary over several orders of magnitude) we consider relative accuracy
and precision; for the other parameters, that set the shape of the
orbit, we consider absolute accuracy and precision.

In this way we obtain for each parameter a distribution that spans all
simulated planetary system; the medians of such distributions can be
taken as the characteristic measure of accuracy or precision for each
parameter. In Tables~\ref{table:accuracy} and \ref{table:precision} we
show the medians of such distributions, along with the 25\% and 75\%
quartiles. For example, the period ($P$) has a relative precision
smaller than $\pm 0.41\%$ in 25\% of the cases in the 5~yr mission.

The accuracy and precision distributions are
presented in Fig.~\ref{fig:errorBarSize}, where they are shown in the
form of a violin plot. In this kind of plot, a
kernel density estimate is calculated for each parameter distribution,
and plotted twice, to form a symmetric figure with normalised area. At
every value of the abscissa, the figure height is proportional to the
density of the distribution.

\subsection{Independence from sky position}
\label{sec:skyposition}
We  assume in Sect.~\ref{sec:astrometricparms} that the
star-planet systems are uniformly distributed on the celestial
sphere. In principle, this assumption might not necessarily be
true. We have found no dependence of either the detection rate or of
any aspect of parameter recovery on the sky position; therefore, no
position bias should be present in our results.

\section{Conclusions}
\label{sec:conclusions}

We have presented an investigation of the planet detection
capabilities of Gaia, using the most up-to-date in-flight properties
of the spacecraft. We have tested a simple Bayesian model of
astrometric data of planetary orbits that relies on off-the-shelf
software and standard procedures for inference and model selection.

We have considered two models: one accounting only for astrometric
parameters (position, parallax, proper motion), and another accounting
for both astrometric and orbital parameters. We investigated model
selection using three information criteria (AIC, BIC, WAIC) and the
$\dchisq$.

Using simulations of Gaia observations of stars with no planets (one
set for a nominal 5~yr mission, and another set for an extended 10~yr
mission), we have compared the information criteria against the
threshold of $\Dic=20$ suggested for strong evidence in favour of one
model over another by Jeffreys' scale for Bayesian evidence. We have
found that the  use of AIC or WAIC with the above threshold allows
false detection fractions that are always $\lesssim 0.6\%$, with AIC being
somewhat stricter than WAIC. The BIC criterion, that puts a stronger
penalty on the more complex model, allowed no false positive out of
6644 cases (5~yr) and 4402 cases (10~yr). The $\dchisq$ can also be
used, but the non-linearity of the model prevents the use of a
reference $\chi^2$ distribution to obtain a p-value; therefore the
choice of a $\dchisq$ threshold needs an empirical calibration.

We have investigated parameter recovery using simulations of stars
with one planet, whose S/N and periods were assigned according to a
logarithmically spaced grid, and whose eccentricities were assigned on
a linear grid. All other parameters are assigned randomly. Realistic
scanning laws are calculated according to a randomly assigned position
on the sky.

We have applied our detection method to 4968 and 4706 simulations of
stars with one planet over a mission of 5 and 10 yr, respectively. We
have chosen a detection threshold of $\Delta\mathrm{BIC}=20$ (i.e.\
with BIC at the strong evidence point according to Jeffreys' scale)
which is equivalent to $\Delta\mathrm{AIC}=70$,
$\Delta\mathrm{BIC}=20$, $\dchisq=80$. This not only minimises the
number of false positives, but has also guaranteed that the detected
systems show a good quality of fit.

We have confirmed the finding of previous studies that the S/N is the
main parameter that determines whether a planet can be detected.  The
50\% detection threshold occurs at $\SN=2.3$ and 1.7 for the 5~yr and
10~yr mission, respectively. Therefore, some approximate maximum
distances to which planets around a 1~$M_\sun$ star can be detected
with a 50\% chance are 39~pc (a Jupiter-mass planet orbiting at 3 au;
5~yr mission), 70~pc (Jupiter-mass, 4 au, 10~yr), 1.9~pc
(Neptune-mass, 3 au, 5~yr), and 3.5~pc (Neptune-mass, 4 au, 10~yr).

The orbital parameters of planets with periods up to the mission
length can be recovered with different degrees of accuracy and
precision. We define accuracy as the difference between the input
value and the median of the output posterior distribution, and  precision
as the length of the 68.3\% HPDI.  Considering the period, in the case
of a 5~yr mission we find a median relative accuracy of $0.70\%$ and a
median relative precision of 1.4\%. For the angular  size of
the semi-major axis of the orbit, we find 7.9\% and 23\%,
respectively.  For the eccentricity, we find 0.077 and 0.19,
respectively.

A follow-up of this work should deal with the detectability of
multiple planets in the same systems, and of very long-period planets,
and should consider data from both Gaia and a future astrometric
mission.

\begin{acknowledgements}
  We thank an anonymous referee whose comments have greatly
  contributed to improve this paper. We thank Martin Gustavsson for
  providing an initial version of our simulation code, and Alessandro
  Sozzetti for useful discussions. We gratefully acknowledge support
  from the Swedish National Space Board (SNSB Dnr 183/14 and SNSB Dnr
  74/14) and the Royal Physiographic Society in Lund.
\end{acknowledgements}

\bibliographystyle{aa}
\bibliography{../fullbiblio}

\end{document}